# REINFORCEMENT OF VACCINE MANDATES AND PUBLIC ATTITUDES TOWARDS VACCINES: WHAT CAN WE LEARN FROM GOOGLE SEARCH ACTIVITY ?

Florian Cafiero* (1,2)
Jeremy K.Ward (3,4)
1- Sciences Po Medialab - 84 rue de Grenelle, Paris, France
2- Columbia University - INCITE - 3078 Broadway, New York, NY, USA
3- CERMES3 (Inserm, CNRS, Ehess, Université De Paris), Villejuif, France
4 - VITROME (Aix Marseille Université, IRD, AP-HM, SSA), Marseille, France

*Correspondence: florian.cafiero@columbia.edu

**Highlights:**

- **We study vaccine-related Google searches in five countries and two American states (2011-2019) .**
- **Vaccine mandates extensions affect attitudes very differently in each country or state.**
- **They very rarely attenuate concerns towards vaccines potential risks.**

**Abstract**:

*International public health policies increasingly favor mandatory immunization. If its short-term effects on vaccine coverage are well documented, there has been little consideration to its effects on public attitudes towards vaccines. In this paper, we examine Google searches related to vaccines in five countries (Australia, France, Germany, Italy, Serbia) and two American states (California) which experienced at least one vaccine mandate extension in the past decade. We found that the effects of a new mandate implementation heavily depends on the context in each specific country or state. We also observed that there is little indication that the passing of new or extended mandates attenuated public doubt towards vaccines.*

*Keywords: Policy ; Vaccines ; Mandatory vaccination; International comparisons; Internet*

In 2019, the World Health Organization put vaccine hesitancy - i.e. “delay in acceptance or refusal of vaccination despite availability of vaccination services” (MacDonald, 2015)-in its list of the ten greatest challenges to global health (World Health Organization (WHO), 2019). In the past decade, awareness of the issues posed by refusal to vaccinate and low immunisation rates more generally has increased globally. The occurrence of several major outbreaks of measles in North America and Europe demonstrated, before COVID-19, the very real threat that infectious diseases still pose even for the wealthiest countries.

Faced with the difficulties of reaching or persuading non-vaccinating parents, public health authorities are increasingly placing their hopes in mandates (Omer et al., 2019). Mandatory vaccination can take many forms (Attwell and Navin, 2019; Harmon et al., 2020). In the past years, several countries and American states have passed new legislation establishing mandatory vaccination or reinforcing pre-existing legislations (Omer et al., 2019). The most commented examples of this trend have been the decision made by the state of California in 2015 to repeal non-medical exemptions to vaccination requirements for school attendance and the announcement made in both France and Italy in 2017 that the list of vaccines required for school admission would be extended (Omer et al., 2019) (Ward, Peretti-Watel et al., 2019). Evidence suggests that mandates are effective in increasing vaccine coverage more often than not (Colgrove, 2016; Greyson et al., 2019; Omer et al., 2019). For instance, studies conducted in the United States have shown that childhood vaccination rates were lower in states in which vaccine exemptions were easier to obtain (Olive et al., 2018; Shaw et al., 2018) and that the states with the most strict mandates seem to be the only ones able to avoid clusters of non-compliance (Garnier et al., 2020). Recent studies attempting to assess the effect of reforms making it more difficult to obtain exemptions in Washington (Omer et al., 2018) and in California (Buttenheim et al., 2018) confirmed the expectation that such measures can help increase vaccine coverage. In France, early assessments of the extension of vaccine mandates show an increase in vaccine coverage even for vaccines not covered by the new mandate (Lévy-Brühl et al., 2019). Recent research on 29 European countries showed that those where vaccination against MMR was mandatory have better vaccine coverage than those without (Vaz et al., 2020). But the introduction or reinforcement of mandates are also known to sometimes backfire and amplify tensions surrounding vaccination (Colgrove, 2016; Omer et al., 2019). Famously, the great antivaccine movements of 19th century Great Britain and early 20th century Brasil emerged in reaction to such measures (Durbach, 2005; Meade, 1986). Historians of vaccination have underlined the fact that the risk of alienating a large part of the public has always been taken into consideration by policymakers (Colgrove, 2006; McCoy, 2019). When negative attitudes towards vaccines are prevalent, it is possible that mandates participate in radicalising hesitants who will resent being forced to vaccinate their children. While this might not appear in vaccine coverage, this can have long-lasting effects on trust in public health institutions and affect coverage of non-mandatory vaccines or new vaccines.

Conscious of these risks, public health authorities in countries with new mandates, such as France (Lévy-Bruhl et al., 2019; Ward et al., 2019), have advocated for a close monitoring of the evolution of attitudes to vaccines. But to date, little research has been devoted to the effect of stricter mandates on attitudes towards vaccines, in today's world where the internet has enabled a rapid sharing of arguments and experience. In Australia, a survey conducted over a year after the “No jab, no pay” reform suggested opposition to vaccines had remained very weak (Smith et al., 2017). In France, a comparison of surveys conducted before and after the mandate extension of 2018 suggests the proportion of vaccine hesitants has decreased but remained at a high level (Cohen et al., 2020; Larson et al., 2018). Research based on survey-type data is essential to assess the proportion of the population susceptible to vaccine hesitancy as well as the social geography of attitudes to vaccines. But the use of such data to assess the evolution of vaccine hesitancy is problematic as it is based on declarations, production cost make it a scarce resource, and the results of each wave of survey can be highly dependent on the short-term context in which it is conducted - i.e. whether vaccines are making the news in the week or month in which the survey was conducted.

In this paper, we wish to contribute to the understanding of the effect of vaccine mandates on attitudes to vaccination by focusing on queries performed on the search engine Google. We assess the effect of the recent reinforcement of vaccine mandates in Australia, France, Germany, Italy, Serbia, California and Washington State, by analysing the evolution of queries performed in each of these countries or states between January 2011 and November 2019. The specificities of each of these policy changes are presented in the next section. Studying the trends of queries on Google presents the advantage of working on people’s spontaneous behavior, and of presenting a continuous stream of data entries, allowing for fine-grained temporal comparisons. For this reason, they are increasingly used as a proxy of attitudes on a variety of subjects ranging from anti-Muslim sentiment (Bail et al., 2018) to beliefs in conspiracy theories (DiGrazia, 2017). In the field of public health, they have been used to help understand the effects of a court decision linking MMR and autism (Aquino et al., 2017), the shift in opinions on vaccination against HPV (Suppli et al., 2018), and the evolution of intentions to purchase chloroquine during the COVID-19 pandemic (Liu et al., 2020)

## MATERIAL AND METHODS

We collected data through the API of Google Trends (Google, 2021). For each country or state, values returned by the API are proportionate to the time and location of a query. Each actual number of searches is divided by the total number of searches in the country or state during the period studied, to get an idea of the relative popularity of the query. The resulting numbers are then scaled on a range of 0 to 100 (Google, 2021).

Our objective was twofold. We aimed to track the evolution of interest in vaccination in general. We therefore monitored the evolution of all queries containing the local words for vaccines or vaccination. We also wished to track interest in the issue of vaccine safety and side effects. To do so, we monitored queries containing both the local words for vaccine and the local words for side effect(s) and danger(s). But, since vaccine hesitancy tends to be focused on some particularly controversial vaccines, we also monitored queries pertaining to the vaccine which focuses doubt in each country or state. In all cases except France, the main controversy seems to have focused on the MMR vaccine and the use of mercury-based adjuvants which are wrongly accused of causing autism. In France, the main debate has been over the use of aluminium-based adjuvants.

We monitored these queries for five countries: Australia, France, Germany, Italy and Serbia ; and two American States: California and Washington. In these six countries, Google is by far the search engine most widely used by Internet users. In our sample of jurisdiction, between 83% (USA) and 97% (Serbia) of queries on the internet were performed on Google in 2019 (Statista, 2019).

We worked on a period ranging from January 1st, 2011 to November, 1st 2019. We have chosen this period which follows the aftermath of the H1N1 influenza crisis, and which precedes the onset of the covid-19 pandemic. Indeed, these two events generated a huge volume of internet queries concerning vaccination. Including these events would completely overshadow all variations observable in recent years. For each country, we highlighted the period between the month of the beginning of a debate around a new vaccine mandate policy (government official announcement, or rumors in a major national media) and its implementation. When no date clearly emerged, we highlighted the six-months period before the policy implementation. For each country or state included in our sample, we present below the context in which the policy change was announced, the nature of the policy change and the timeframe of its implementation.

We detected abnormally high or low levels of Google searches by running a time series outlier detection algorithm (Chen and Liu, 1993). To that end, for each of our datasets, we run an automated search for the best Autoregressive integrated moving average (ARIMA) model - the widest class of models for forecasting a time series. The parameters ($p$,$d$,$q$) - where $p$ is the number of autoregressive terms ; $d$ the number of nonseasonal differences needed for stationarity ; $q$ the number of lagged forecast errors in the prediction equation - are automatically selected to optimize the Akaike Information Criterion. These ARIMA models also help us assess the impact of vaccine mandate extensions on queries in the short and longer run. Detailed results for each of the 21 models are provided in the supplementary file.

To understand which events could reasonably explain why such peaks have occurred, we search on Google the date at which the major peaks occur and the query monitored (e.g. “vaccine august 2011”).

Australia:

Since 1998, the Australian state fines individuals who do not comply with vaccine recommendations for their children, and to condition childcare subsidies to the satisfaction of this requirement (Ward et al., 2013). A “Conscientious Objector Form” was however available for people refusing vaccines. In 2013, a major local newspaper started a campaign named “No Jab, No Pay”, which rapidly gained national interest (Beard et al., 2017). Following a debate at the Australian Parliament, the government announced in April 2015 the end of this exemption system, enforced on January 1st, 2016. (Department of Health Australia, 2015)

France:

France has a century-old history of vaccine mandates - the first occurrence of such a policy dating back to 1902 and the smallpox vaccine. Since then vaccination against diphteria (1938), tetanus (1940), tuberculosis (1950) and polio (1964) have become compulsory (Moulin, 2014). But all vaccines introduced after 1966 were not made mandatory. Later, the mandates for smallpox and tuberculosis vaccination were repealed (in 1984 and 2007 respectively). In June 2017, the newly appointed Health Minister announced in a press interview that the government would make all 11 recommended childhood vaccines mandatory to access childcare and schools (Mari et al., 2017). This was confirmed by the Prime Minister on July 5th. The law has been enforced since the 1st of January 2018 (Ward et al., 2018).

Germany

After Germany's reunification in 1990, vaccines, which were mandatory only in the East, became voluntary across the country. Following some doubts in the public regarding vaccine safety, as well as measles outbreaks, the German government passed in July 2015 a National Preventive Healthcare Act (Willmann, 2018), forcing parents to show evidence that they discussed the subject of vaccination with a medical practitioner if they wanted their child to attend daycare. Since July 2017, kindergartens are legally binded to alert public authorities if parents have not provided the required evidence. Parents can face a fine of up to 2500$ (Attwell et al., 2018).

Italy

As in France, Italy has long had a system where some vaccines were voluntary while others were compulsory (diphtheria since 1939, polio since 1966, tetanus since 1968 and hepatitis B since 1991) (D'Ancona et al., 2018). The year 1999 marked the beginning of a decade of what Carlo Signorelli (Signorelli, 2019) called "the way to informed acceptance" ("*la via dell'adesione consapevole"*). In particular, the Italian government chose to accept in school children who did not receive mandatory vaccines. Yet in the year 2010s, vaccine hesitancy exploded in Italy, which gradually convinced the authorities to deviate from this path. At the beginning of 2017 a National Immunization Prevention Plan ("Piano Nazionale di Prevenzione Vaccinale") was designed, and proposed (Signorelli et al., 2017). In April 2017, the extension of vaccine mandates starts being discussed, and a decree-law issued in June 2017, extending the list of mandatory vaccines from 4 to 10, was finally adopted by the Italian Parliament on July 31st, 2017 (D'Ancona et al., 2018), resulting in a ministerial letter on September 1st, 2017.

Serbia

Fairly uncontroversial during the socialist era (Trifunović, 2019), mandatory vaccines were still positively perceived in the year 2000s, when there were no particular issues regarding childhood immunization (Šterić et al., 2007). Attitudes towards vaccines shifted after the 2009 H1N1 crisis (Trifunović, 2019). In Serbia, immunisation is mandatory against 11 diseases: diphtheria, haemophilus influenzae type B, hepatitis B, measles, mumps, pertussis, poliomyelitis, rubella, tetanus, tuberculosis, and since April 2nd, 2018, pneumococcus (N1, 2018). A law passed in February 2016 states that parents refusing to vaccinate their children can be fined up to 1400 €. Repeated offenses can lead to imprisonment and the launch of a social services investigation for child neglect (Trifunović, 2019).

United States

Each American state requires at least some vaccines for a child to attend daycare or school (Bednarczyk et al., 2019). Vaccine regulations in the United States vary from state to state. Medical exemptions exist in each state, but some states, such as California or Washington, allow non-medical exemptions (religious or philosophical) (Bednarczyk et al., 2019)

California

On January 1st, 2012, California enforced Assembly Bill 2109, requiring parents to prove they consulted a medical professional about vaccines - a new policy which resulted in a reduction of non-medical exemption by 25% (Buttenheim et al., 2018).

A new bill (Senate Bill 277) signed in July 2015 (Mello et al., 2015), and enforced from January 1st 2016 on, eliminated nonmedical exemptions for school entry. This raised concerns about medical exemptions being used instead of nonmedical exemptions (Mohanty, 2018). To prevent this trend from further developing, a new bill, (Senate Bill 276) was examined on April 24th, 2019, adopted in September 2019 and came into effect on January 1st, 2020. From then on, a state health official will have to review all medical exemptions at schools who did not share their vaccination rates, in which less than 95% of students are actually vaccinated, or from doctors who submit five or more exemptions within a year. In case of suspected fraud, physicians will be reported to California's medical board, and exemptions reversed.

Washington

At the beginning of the 2010s, Washington state had one of the highest rates of nonmedical exemptions in the USA (Constable et al., 2014). Faced with a pertussis outbreak in 2010, the state changed its exemption law in July 2011, requiring parents looking for a non-medical exemption to prove they discussed the risks and benefits of vaccines with a medical practitioner. After debates in 2015 about tightening these rules (Corte, 2015), a measles outbreak sickening 74 persons in the state in 2019 prompted a new reform, passed by the Washington senate in April 2019, abrogating philosophical exemptions for the MMR vaccine (Gutman, 2019).

**RESULTS**

First of all, vaccination policy reforms are of course not the only driving forces behind the evolution of Internet queries on vaccination. It is remarkable that vaccine mandate extensions are very rarely the moments around which Google searches around vaccination increase the most (Figure 1). The example of the H1N1 flu pandemic is exemplary in this regard. In Australia, a southern hemisphere country, an H1N1 vaccine in October/November was simply not a major issue. However, in all other countries in our database except Italy, searches at the time of the H1N1 pandemic widely outnumber the peaks of the following ten years. Searches during this pandemic ranged from almost twice as many as during the peak of Google searches during a vaccine mandate extension in California or the Washington State, to more than 14 times as many in France. Other important peaks of interest are also linked to epidemic episodes. The most important peaks during the 2011-2019 period occurred during measles outbreaks (in Serbia, California and Washington State) or severe flu epidemics (in Australia). Only in Italy did the reinforcement of mandates have a greater impact on Google queries on vaccination than any epidemics.

Figure 2 shows the evolution of Google searches on vaccination and vaccination hazards in our seven geographical regions of interest. Figure 3 focuses only on queries related to the "dangers" of vaccines, for better readability. As we can see, some trends are common to all or most of the countries and states studied here. The seasonality of searches varies from country to country, but is almost always present. This shows the important impact of school calendars and flu season on the questions that Internet users ask themselves about vaccination. However, the amplitude of this seasonality differs greatly. It is extremely strong in a country such as Australia, for example. In fact, the flu season is sometimes a time when Google searches reach their peak over the 2011-2019 period, as in France or Germany (fig. 1).

Yet, each time a discussion about a new vaccine mandate takes place, the search volumes regarding both vaccination in general, and their alleged dangers, increase significantly, if not dramatically. The only exception is Serbia, where almost no impact is observed. But the discussion of a policy change does not seem to positively or negatively influence the trend in Google searches. The medium-term effects of these decisions on interest in vaccination, or concern about its alleged side effects, seem minor or imperceptible. An exception to this observation would appear to be Italy, where searches on these themes have even decreased after the decision to extend the vaccination obligation. There is also a temporary change in trend in California. For a year and a half, searches on vaccines stagnated rather than increased - but eventually the trend began to pick up again. Finally, in France, searches about the risks of aluminum in vaccines increased during the debates preceding the mandate extension, but decreased after it was passed. On the other hand, queries about the risks of vaccines slightly increased after the 2017 reform.

We can also observe major differences between countries. It is striking to note how the proportion of vaccine related queries that focus on concerns about the risks of vaccines varies from country to country. In particular, the percentage of such queries in France and Germany is much greater than in any other country or state in our sample.

| | H1N1: peak of queries | H1N1: date of the peak | Vaccine mandate extension: peak of queries | Mandate extension: date of the peak | 2011-2019: global peak of queries: | Possible reason of the peak |
|---|---|---|---|---|---|---|
| Australia | 54 | Oct, 2009 | 80 | Apr, 2015 | 100 | Flu epidemics-07/2019 |
| France | 100 | Nov, 2009 | 7 | Jul, 2017 | 7 | Flu vaccination campaign - shot by pharmacists for the first time - 10/2019 |
| Germany | 100 | Nov, 2009 | 19 | Jul, 2015 | 19 | Flu vaccination campaign - 10/2019 |
| Italy | 88 | Nov, 2009 | 100 | Sep, 2017 | 100 | Vaccine mandate extension - 09/2017 |
| Serbia | 100 | Nov, 2009 | 24 | Mar, 2016 | 40 | Measles outbreak - 02/2018 |
| USA - California | 100 | Oct, 2009 | 42 | Sep, 2019 | 56 | Measles outbreak - 02/2015 |
| USA-Washington State | 100 | Oct, 2009 | 36 | Apr, 2019 | 54 | Measles outbreak - 01/2019 |

*Figure 1 - Heatmap: extreme values of vaccine-related queries (as a global topic) on Google from January 2004 to November 2019 : peak of queries during the H1N1 pandemics ; peak of queries during the vaccine mandate extension generating the most searches during the period ; absolute peak of queries from January 2011 to November 2019. 100 denotes the most important value during the 2004-2019 period.*

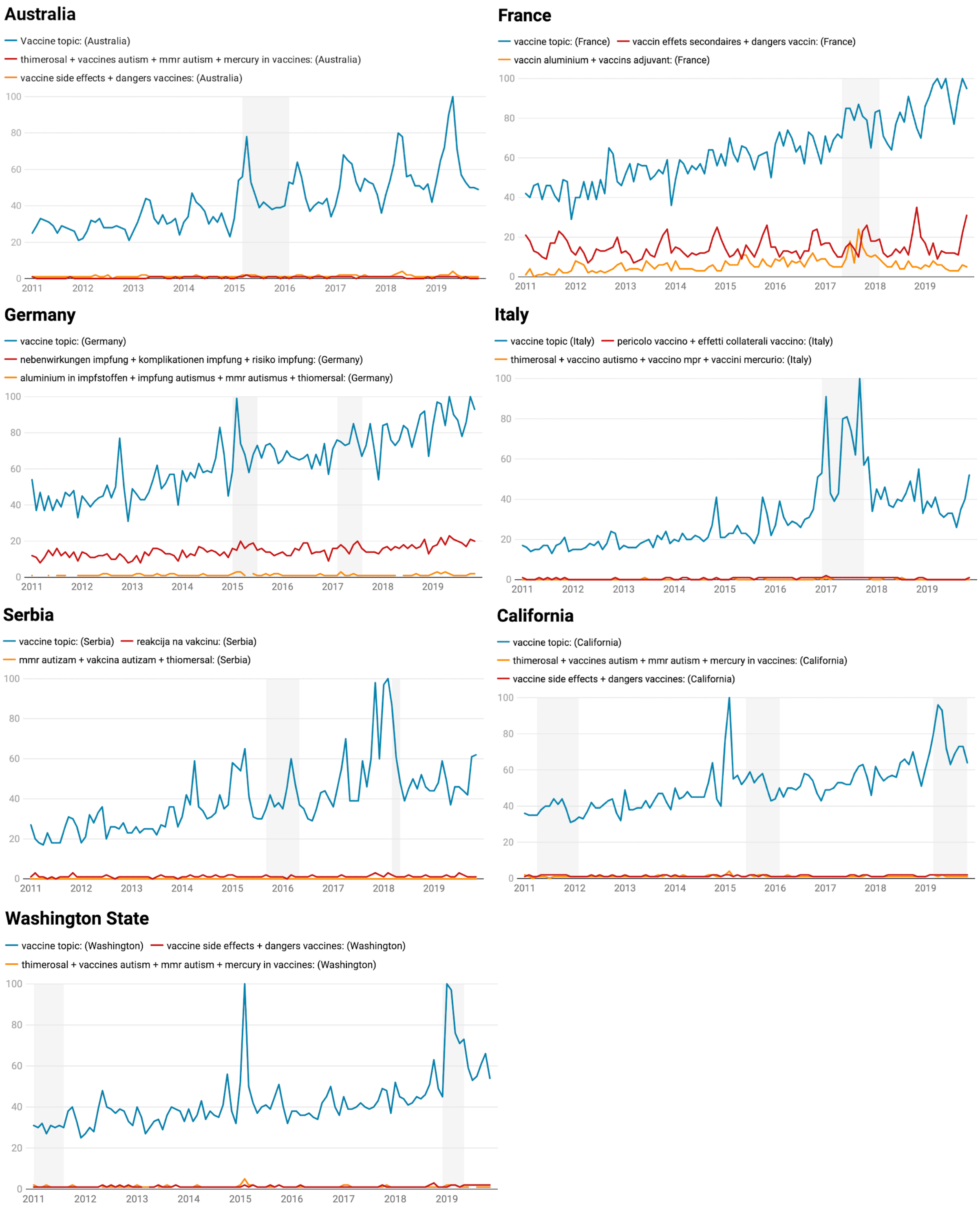

*Figure 2 - Evolution of Google queries for vaccine-related keywords from January 2011 to November 2019: in blue, keywords related to the global vaccine topic ; in red, keywords related to the global "dangers" of vaccines ; in yellow, keywords related to specific vaccine controversies.*

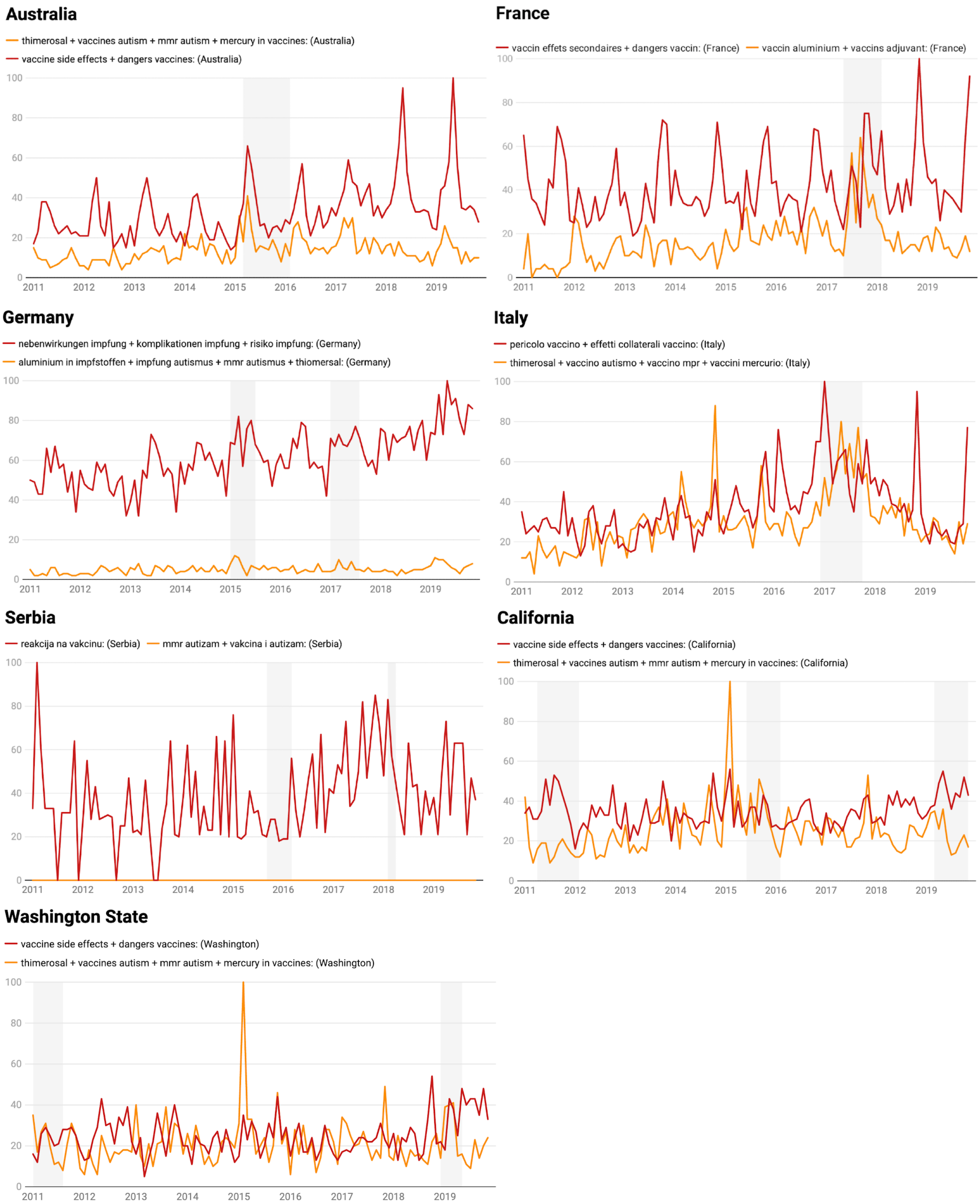

*Figure 3 - Evolution of Google queries for vaccine-related keywords from January 2011 to November 2019: in red, keywords related to the global "dangers" of vaccines ; in yellow, keywords related to specific vaccine controversies ; in grey, period of discussion then enforcement of the new mandates.*

| Geographical area | Peak(s) of general queries about vaccines | Peak(s) of queries about vaccines' risks (general) | Peak(s) of queries about vaccines' risks (specific) |
|---|---|---|---|
| Australia | Yes | Yes | Yes |
| France | No | Yes | Yes |
| Germany | Yes in 2015<br>No in 2017 | No | Yes in 2015<br>No in 2017 |
| Italy | Yes | Yes | Yes |
| Serbia | No | No | No |
| California | No in 2011<br>No in 2015<br>Yes in 2019 | No in 2011<br>No in 2015<br>Yes in 2019 | No |
| Washington | No in 2011<br>Yes in 2019 | Yes | Yes |

*Figure 4 - Synthesis of the Chen and Liu's peak detection procedure results on the temporal series of Google searches: presence or absence of a statistically significant peak during a vaccine mandate extension (see supplementary material).*

## DISCUSSION

We analysed the evolution of vaccine-related queries on google performed between January 2011 and November 2019 in seven countries or American States which passed new more strict legislation on vaccination during that period (Australia, France, Germany, Serbia, Italy, the State of California and the State of Washington). Our main finding is that the passing of these legislations was rarely associated with a decrease of hesitancy-driven queries. The only exceptions are Italy and, more marginally and for a limited period of time, California. On the other hand, in Australia and Serbia, the passing of new mandates was followed by an increase of queries on vaccination in general and side effects in particular. We also found that the period between the announcement of the new legislation and its application was associated with increased interest in vaccination, including more queries directly pertaining to the purported side effects of vaccines.

Using Google Trends data induces various limitations. It is, of course, difficult to establish with certainty the link between online research and actual opinion or behavior (Bail, 2014; Golder and Macy, 2014). For example, some Internet users looking for details on a theory about the dangers of vaccination might not really be worried about it, but simply want to know more about a theory they have heard but consider to be far-fetched, or want to criticize. It seems unlikely, however, that this would be the prevailing pattern. Cultural variations in the use of the Internet in general, and of the search engine Google in particular, could also bias our study, even if, as we have shown above, its use is extremely widespread in each of the countries studied. However, while vaccine controversies have been much studied on social networks, they have been much less studied in relation to queries on Google, even though this is one of the main ways people look for information on the Internet. Future studies could combine these two perspectives to better understand the evolution of attitudes towards vaccination.

Our results have several implications for public health.

The main implication is that the effect of mandates on attitudes depends greatly on national contexts. Public authorities have historically been somewhat aware of the complex effects of mandates (Colgorove, 2006; McCoy, 2019). This is testified in the great variety of forms taken by mandatory vaccination around the globe which tend to be adapted to the precise issues faced at the moment of their design and the local institutional idiosyncrasies (Attwell and Navin, 2019; McCoy, 2019). But in the current context where public health authorities increasingly (rightly) rely on international academic publications and reports by trans-national organisations to devise their policy, there is a risk of relying too heavily on the experiences of a limited number of countries and adopt a “one size fits all” approach. For this reason, most specialists of vaccination behavior recommend great caution and attention to local contexts when deciding whether and how to mandate vaccines (Omer et al., 2019).

Secondly, in most countries and states we studied, there is no indication that the passing of more coercive legislation attenuated public doubt towards vaccines. This implies that public health experts should be weary of not relying too heavily on vaccine coverage data to assess the multifaceted effects of a mandate. Coverage determines the actual circulation of viruses, yes. But, people’s adherence to non-mandatory and future vaccines is not independent from their prior experience of the more common (and potentially mandatory) vaccines. So does their trust in the health system overall. France is a good case in point. Data from the past ten years suggests it is one of the most vaccine-hesitant countries in the world (Ward, Peretti-Watel et al., 2019), with a wide network of actors criticizing vaccines on the internet and in the media (Cafiero et al., 2021). Following the new legislation of 2018, coverage rates increased dramatically but negative attitudes to vaccines only slightly decreased (Lévy-Brühl et al., 2019). Recent studies suggest that this persistent high level of vaccine hesitancy and low trust in public health authorities is likely to pose a problem when a COVID-19 vaccine will be released. In April, at the height of the epidemic in France, a quarter of respondents to several large surveys said they would not get vaccinated against this virus (Ward et al., 2020). In the case of COVID-19, the international public health community seems very aware of the necessity to make efforts to build trust for this vaccine even if vaccination becomes mandatory (Mello et al., 2020). Our results suggest that most specialists of mandatory vaccination are right to emphasize the necessity to always complement vaccine mandates with other forms of intervention targeted at building trust in vaccines (Attwell and Navin, 2019).

Communication is all the more important that policy changes do not seem to constitute the main driver of internet queries in our study. The announcement, discussion or application of new mandatory vaccination policies seem to be only one of several events susceptible to increase interest in vaccination and their purported side-effects in at least some countries. In our case studies, we found spikes in number of internet searches associated with epidemic outbreaks (such as the measles outbreak of 2015 but also the 2009 H1N1 influenza pandemic and the current COVID-19 outbreak), the launch of the seasonal flu vaccination campaign, and the annual procedure of school enrolment for children (in Australia). These very different types of events have in common the fact that they tend to be very widely covered by the media. Much like these other events, the passing of new legislation seems to put vaccines under the collective looking glass. This is important because experts in vaccination behaviours have emphasized the importance of integrating vaccination in the daily routine of people, and parents in particular. To reach wide coverage, vaccination must become a normal, almost automatic act, one that is not constantly questioned (Dubé et al., 2018). Indeed, belief in the safety and efficacy of recommended vaccines is regularly compromised by the irruption of vaccine-critical communication in the daily lives of people via multiple communication channels: relatives and friends, content recommendations on social media, media debates… Parents might not be directly convinced by such discourses, but they nonetheless can have the effect of foregrounding vaccines as an object of concern, necessitating further investigation, which can in turn lead to more consultation of vaccine-critical information easily accessible via a simple query on Google or elsewhere on the internet (Peretti‐Watel et al., 2019; Sobo et al., 2016). Specialists of the ethics of vaccination have often underlined this paradox: in a context where disinformation is everywhere and where people's cognition can fall prey to a number of cognitive biases, the search for more information can be detrimental to adherence to vaccination (Navin, 2015). If the passing of new legislation leads to further interest in vaccines and their side effects, as we have found, then it is crucial that the public and journalists can easily find reliable information on vaccines in the fewest number of clicks possible. Our data suggests this information must be available at the moment of the announcement of the upcoming change in legislation, not later, when it is finally put into effect. This means that communication efforts must not only accompany the passing of stricter mandates, it must precede it. The effectiveness of mandates paradoxically rests on the early application of the measures that can potentially render the recourse to mandates superfluous.

Whether the path leads to mandatory vaccination or not, such as in the case of a future COVID-19 vaccine, our results suggest that the first step is always building trust and providing people, healthcare providers and journalists with reliable and easily accessible information.

# Outlier detection

## Germany

```
## Registered S3 method overwritten by 'quantmod':
##   method            from
##   as.zoo.data.frame zoo

## Series: allem.ts
## Regression with ARIMA(0,1,1) errors
##
## Coefficients:
##           ma1     AO22     AO50      AO84
##       -0.7467  31.2741  34.4616  -23.7782
## s.e.   0.0525   6.9885   6.9881    6.9890
##
## sigma^2 estimated as 58.11:  log likelihood=-364.08
## AIC=738.16   AICc=738.76   BIC=751.47
##
## Outliers:
##   type ind time coefhat  tstat
## 1   AO  22   22   31.27  4.475
## 2   AO  50   50   34.46  4.931
## 3   AO  84   84  -23.78 -3.402
```

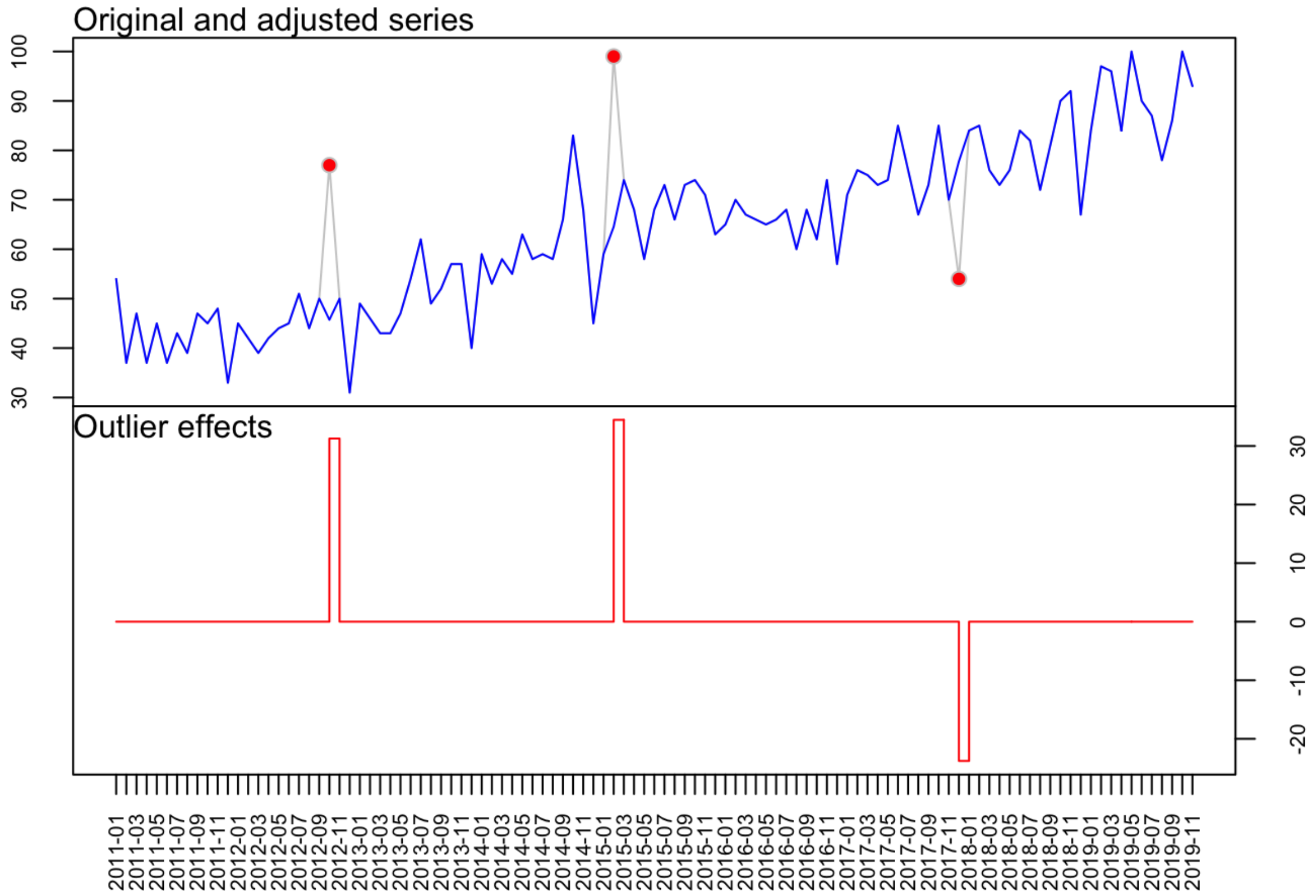

# Australia

```
## Series: aust.ts
## Regression with ARIMA(1,0,0) errors
##
## Coefficients:
##          ar1  intercept     LS50     AO52     TC75     TC88    TC100    AO101
##       0.6598    30.3627  18.5739  23.9490  18.4392  21.7686  25.7533  21.6525
## s.e.  0.0782     2.0965   2.7861   4.4544   5.3820   5.5071   5.6602   4.4769
##
## sigma^2 estimated as 30.63:  log likelihood=-331.04
## AIC=680.08   AICc=681.93   BIC=704.13
##
## Outliers:
##   type ind time coefhat tstat
## 1   LS  50   50   18.57 6.667
## 2   AO  52   52   23.95 5.376
## 3   TC  75   75   18.44 3.426
## 4   TC  88   88   21.77 3.953
## 5   TC 100  100   25.75 4.550
## 6   AO 101  101   21.65 4.837
```

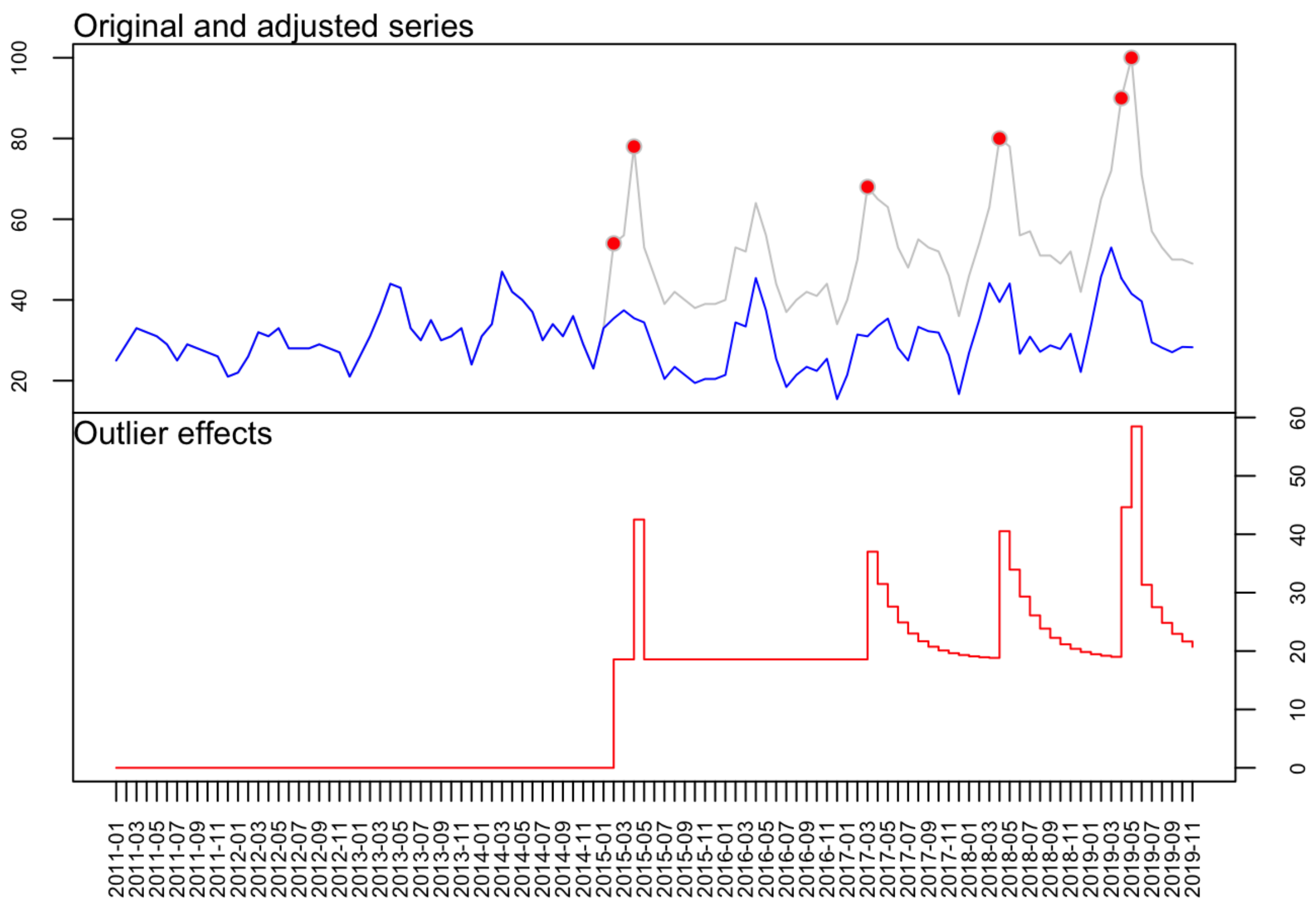

# California

```
## Series: calif.ts
## Regression with ARIMA(0,1,2) errors
```

```
##
## Coefficients:
##          ma1      ma2     AO25     TC49     AO50      AO84     AO99    TC100
##      -0.1114  -0.6234  11.7897  26.6667  37.0153  -11.1758  17.2805  31.4413
## s.e.   0.0945   0.0916   2.8128   4.3104   3.2301    2.8141   4.0590   5.6294
##
## sigma^2 estimated as 21.02:  log likelihood=-308.21
## AIC=634.41   AICc=636.29   BIC=658.38
##
## Outliers:
##   type ind time coefhat  tstat
## 1   AO  25   25   11.79  4.191
## 2   TC  49   49   26.67  6.187
## 3   AO  50   50   37.02 11.459
## 4   AO  84   84  -11.18 -3.971
## 5   AO  99   99   17.28  4.257
## 6   TC 100  100   31.44  5.585
```

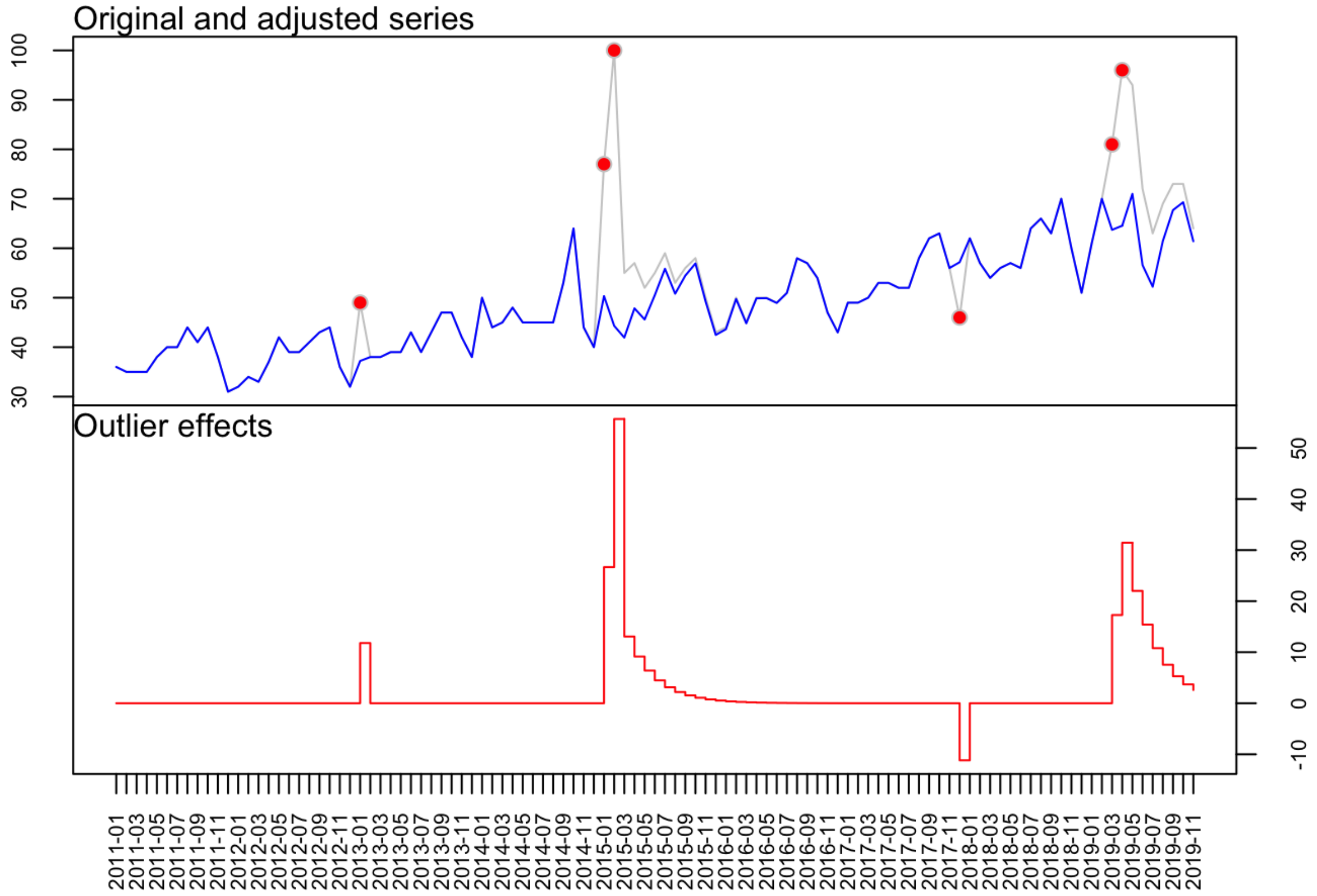

# France

```
## Series:
## ARIMA(0,1,1)
##
## Coefficients:
##          ma1
##      -0.6704
```

```
## s.e.   0.0802
##
## sigma^2 estimated as 57.46:  log likelihood=-364.91
## AIC=733.83   AICc=733.94   BIC=739.15
##
## No outliers were detected.
```

```
## 'x' does not contain outliers to display
```

```
## NULL
```

# Italy

```
## Series: ita.ts
## Regression with ARIMA(1,1,1) errors
##
## Coefficients:
##          ar1      ma1     AO47     AO58     LS71     AO73     TC77     AO81
##       0.4225  -0.8564  17.3939  13.5810  16.7709  43.6383  39.8833  42.5481
## s.e.  0.1391   0.0643   3.7664   3.7714   3.3665   3.7916   4.5143   3.8269
##          AO83     AO95
##       15.1395  18.1873
## s.e.   3.7818   3.7756
##
## sigma^2 estimated as 20.81:  log likelihood=-306.34
## AIC=634.68   AICc=637.49   BIC=663.98
##
## Outliers:
##   type ind time coefhat  tstat
## 1   AO  47   47   17.39  4.618
## 2   AO  58   58   13.58  3.601
## 3   LS  71   71   16.77  4.982
## 4   AO  73   73   43.64 11.509
## 5   TC  77   77   39.88  8.835
## 6   AO  81   81   42.55 11.118
## 7   AO  83   83   15.14  4.003
## 8   AO  95   95   18.19  4.817
```

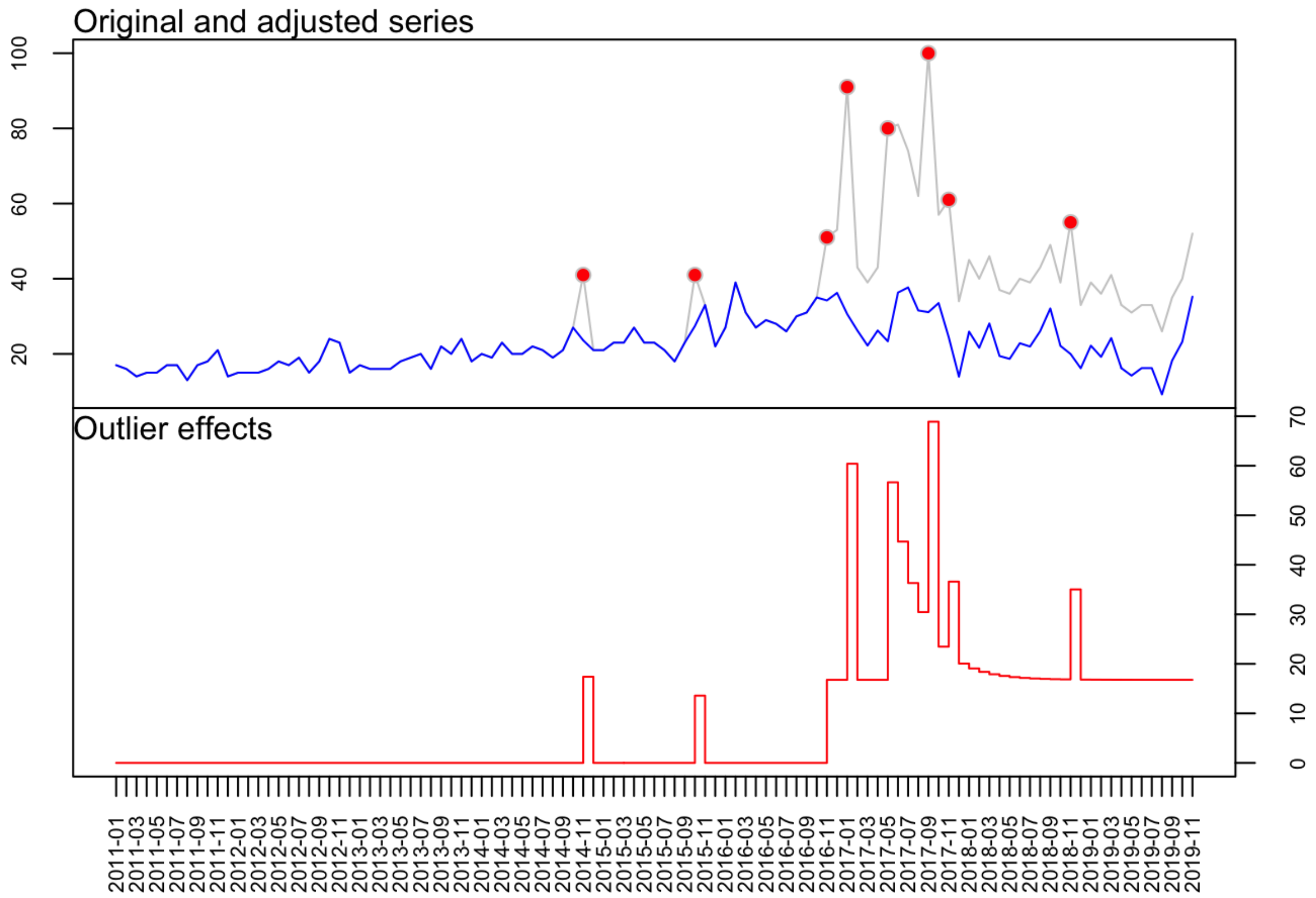

# Serbia

```
## Series: serb.ts
## Regression with ARIMA(0,1,3) errors
##
## Coefficients:
##          ma1      ma2      ma3     AO40     TC49     AO83     TC85
##      -0.3072  -0.2509  -0.2325  23.0268  25.0210  41.6163  43.2382
## s.e.   0.1069   0.0986   0.1019   5.9750   7.1771   6.7567   7.4504
##
## sigma^2 estimated as 56.52:  log likelihood=-361.03
## AIC=738.07   AICc=739.55   BIC=759.37
##
## Outliers:
##   type ind time coefhat tstat
## 1   AO  40   40   23.03 3.854
## 2   TC  49   49   25.02 3.486
## 3   AO  83   83   41.62 6.159
## 4   TC  85   85   43.24 5.803
```

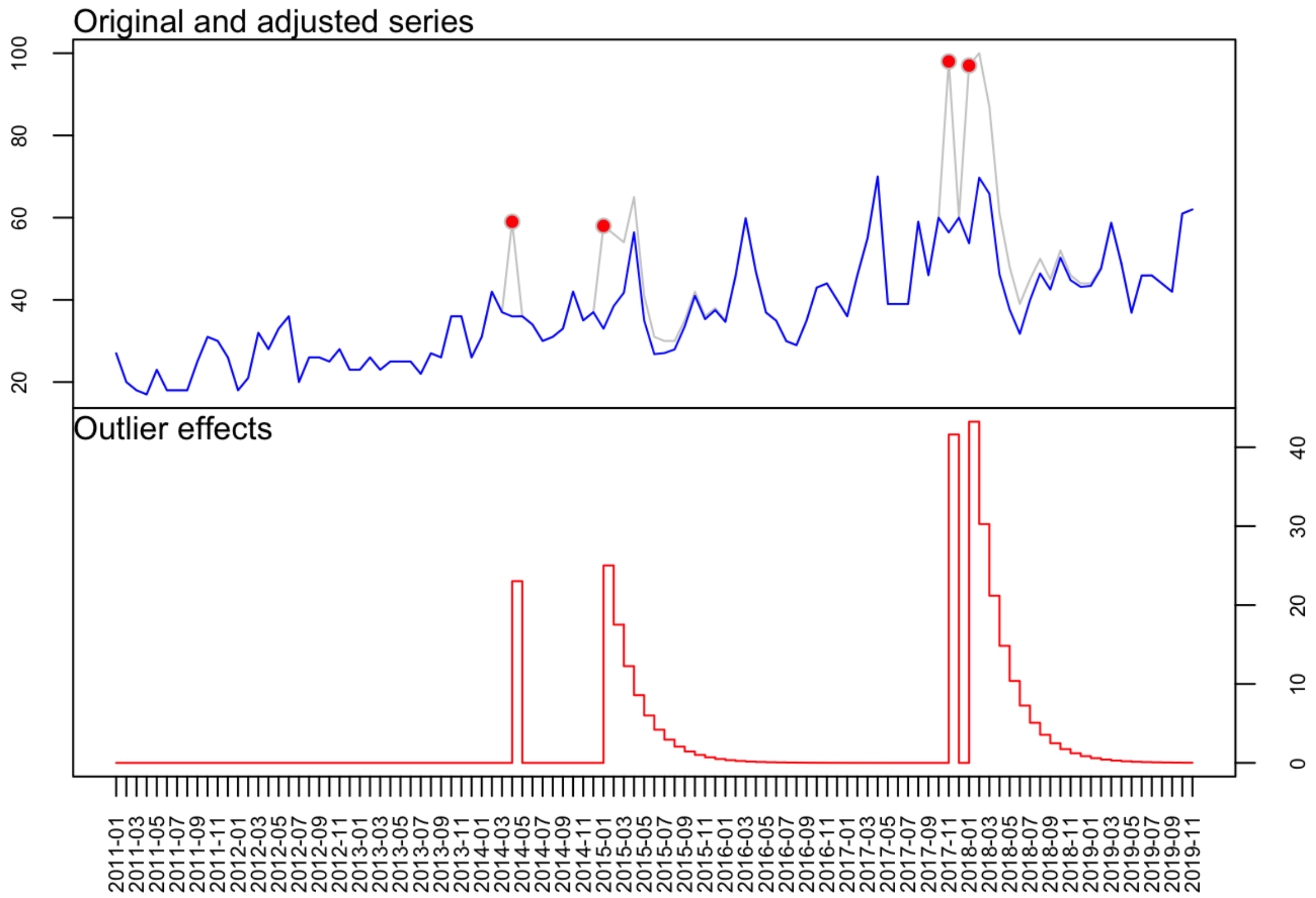

# Washington

```
## Warning in locate.outliers.iloop(resid = resid, pars = pars, cval = cval, :
## stopped when 'maxit.iloop' was reached

## Series: wash.ts
## Regression with ARIMA(0,0,1) errors
##
## Coefficients:
##          ma1  intercept    AO46    LS49     AO50     TC85     LS94     TC97
##       0.6899    34.4491  13.0005  6.9803  45.5040  15.0207  13.8264  44.9688
## s.e.  0.1183     1.0706   3.4580  1.5401   3.7052   4.6407   2.3440   4.7140
##
## sigma^2 estimated as 21.13:  log likelihood=-311.19
## AIC=640.39   AICc=642.24   BIC=664.44
##
## Outliers:
##   type ind time coefhat  tstat
## 1   AO  46   46   13.00  3.759
## 2   LS  49   49    6.98  4.532
## 3   AO  50   50   45.50 12.281
## 4   TC  85   85   15.02  3.237
## 5   LS  94   94   13.83  5.899
## 6   TC  97   97   44.97  9.539
```

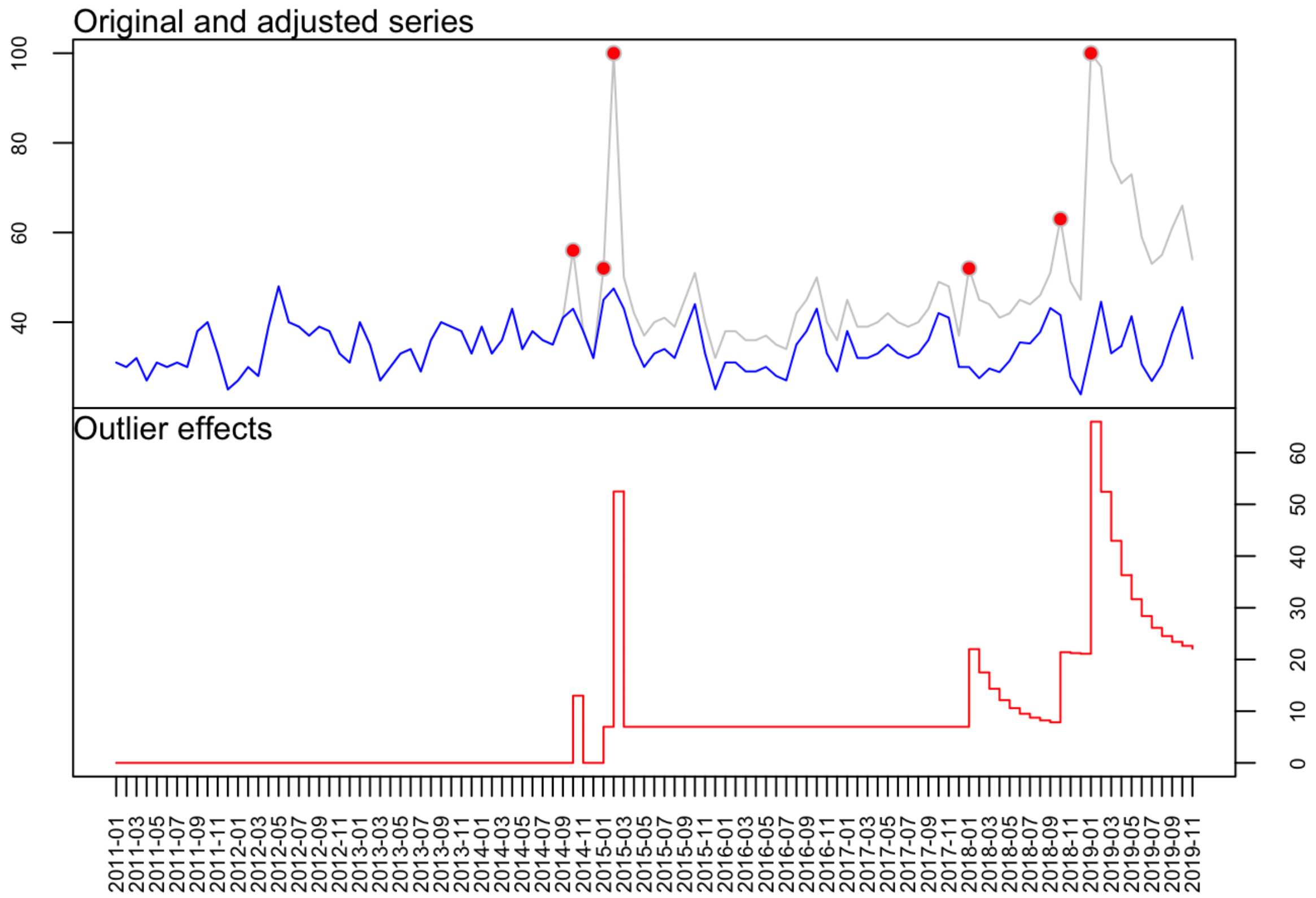

# Outlier detection: risks

## Germany (risks)

### General

```
## Series:
## ARIMA(0,1,1)
##
## Coefficients:
##           ma1
##       -0.7240
## s.e.   0.1241
##
## sigma^2 estimated as 102:  log likelihood=-395.4
## AIC=794.8   AICc=794.92   BIC=800.13
##
## No outliers were detected.
```

```
## 'x' does not contain outliers to display
```

```
## NULL
```

### Specific

```
## Series: allem_risk_spec.ts
```

```
## Regression with ARIMA(0,1,1) errors
##
## Coefficients:
##           ma1    TC49    TC98
##       -0.9259  5.8845  6.6959
## s.e.   0.0416  1.3305  1.3817
##
## sigma^2 estimated as 3.126:  log likelihood=-210.27
## AIC=428.54   AICc=428.94   BIC=439.2
##
## Outliers:
##   type ind time coefhat tstat
## 1   TC  49   49   5.885 4.423
## 2   TC  98   98   6.696 4.846
```

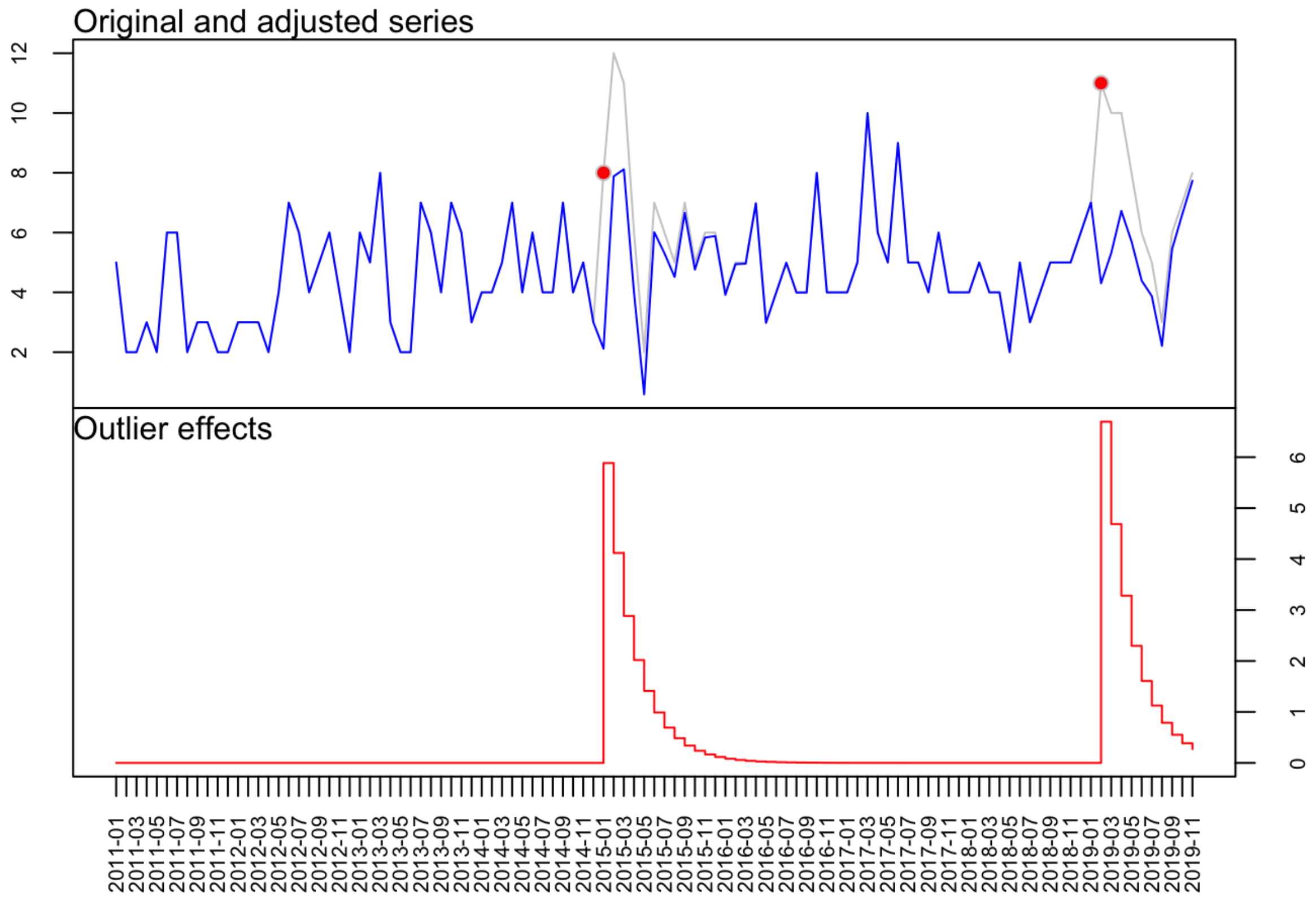

# Australia (risks)

## General

```
## Series: aust_risk_gen.ts
## Regression with ARIMA(0,1,1) errors
##
## Coefficients:
##           ma1     TC37     TC50     AO52     TC63     TC75     TC99
##       -0.8712  10.0120  16.8340  20.6869  14.3564  17.0754  12.7542
## s.e.   0.0548   2.8807   2.9988   3.6534   2.9130   3.0528   3.1120
```

```
##
## sigma^2 estimated as 14.15:  log likelihood=-287.95
## AIC=591.89   AICc=593.37   BIC=613.2
##
## Outliers:
##   type ind time coefhat tstat
## 1   TC  37   37   10.01 3.476
## 2   TC  50   50   16.83 5.613
## 3   AO  52   52   20.69 5.662
## 4   TC  63   63   14.36 4.928
## 5   TC  75   75   17.08 5.593
## 6   TC  99   99   12.75 4.098
```

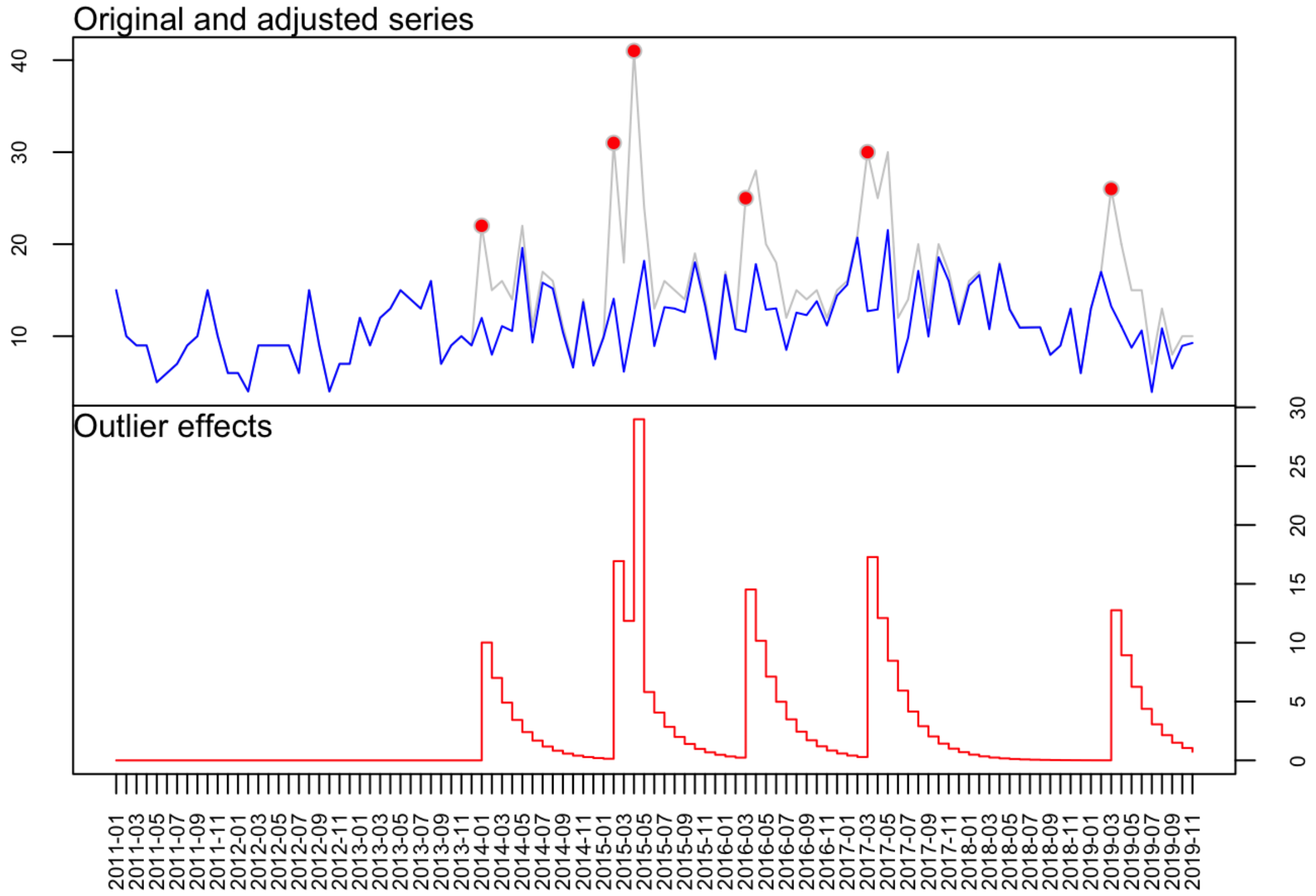

## Specific

```
## Warning in log(s2): production de NaN
```

```
## Series: aust_risk_spec.ts
## Regression with ARIMA(0,0,1) errors
##
## Coefficients:
##          ma1  intercept     TC52     LS76     AO89    AO101
##       0.6690    27.2876  31.2099  14.2877  33.8188  35.6379
## s.e.  0.0869     1.5424   7.1304   2.7471   6.1360   6.9259
##
## sigma^2 estimated as 66.4:  log likelihood=-373.51
## AIC=761.02   AICc=762.15   BIC=779.73
```

```
##
## Outliers:
##   type ind time coefhat tstat
## 1   TC  52   52   31.21 4.377
## 2   LS  76   76   14.29 5.201
## 3   AO  89   89   33.82 5.512
## 4   AO 101  101   35.64 5.146
```

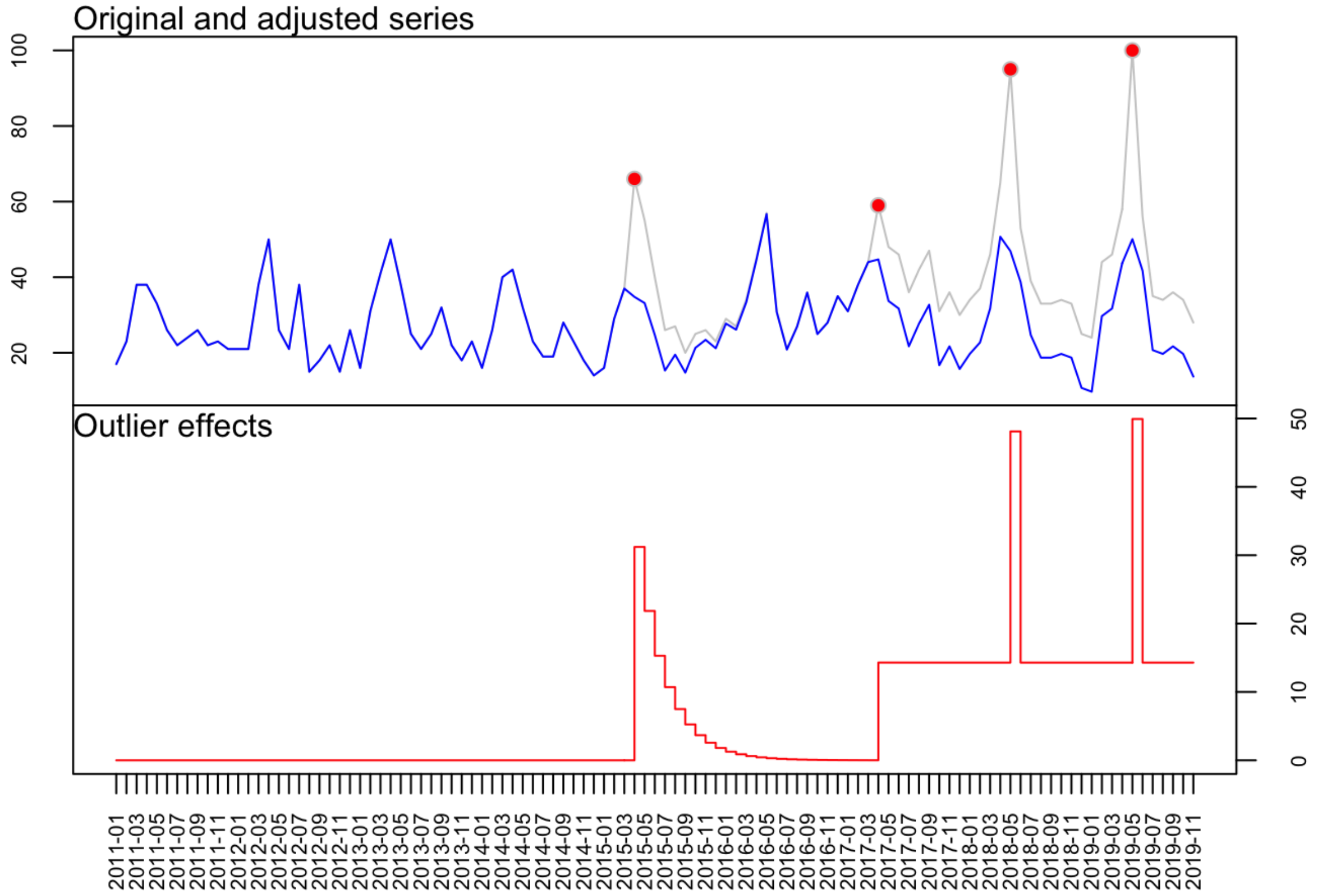

# California (risks)

## General

```
## Series: calif_risk_gen.ts
## Regression with ARIMA(1,0,0) errors
##
## Coefficients:
##          ar1  intercept     AO46     AO50    LS100
##       0.2859    33.3742  20.8235  22.0306  12.1668
## s.e.  0.0922     0.9272   6.3492   6.3512   3.2034
##
## sigma^2 estimated as 45.52:  log likelihood=-353.59
## AIC=719.17   AICc=720.01   BIC=735.21
##
## Outliers:
##   type ind time coefhat tstat
## 1   AO  46   46   20.82 3.280
```

```
## 2   AO  50    50    22.03 3.469
## 3   LS 100  100    12.17 3.798
```

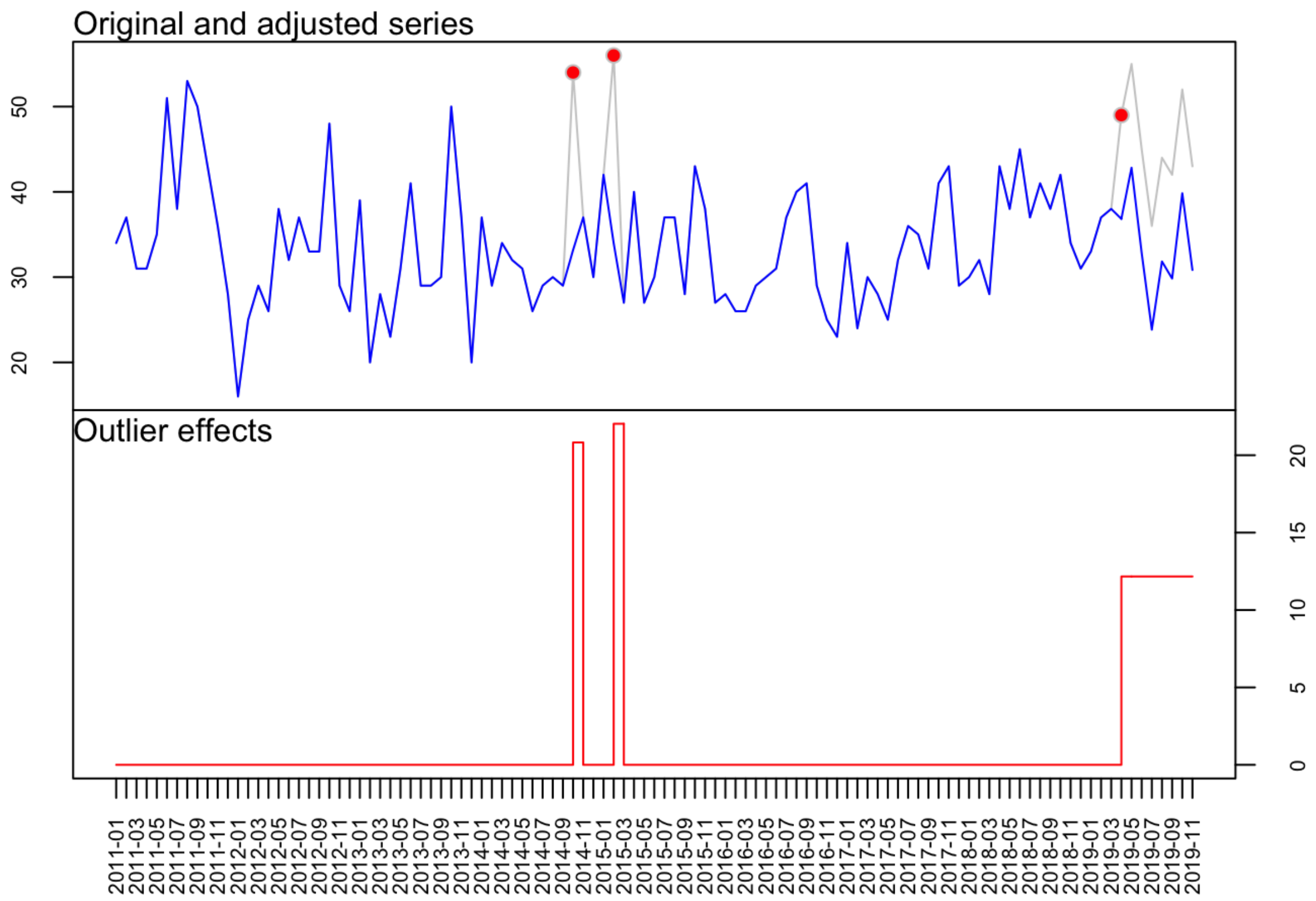

## Specific

```
## Series: calif_risk_spec.ts
## Regression with ARIMA(1,0,0) errors
##
## Coefficients:
##          ar1  intercept      LS2     TC49     AO50     AO55     TC57     AO83
##       0.4501    44.2869  -22.2067  33.8526  57.8159  20.0498  25.7242  28.7343
## s.e.  0.0872     6.8720    6.8877   6.6369   6.3181   6.2491   6.5853   6.2478
##
## sigma^2 estimated as 50.52:  log likelihood=-357.63
## AIC=733.27   AICc=735.12   BIC=757.32
##
## Outliers:
##   type ind time coefhat  tstat
## 1   LS   2    2  -22.21 -3.224
## 2   TC  49   49   33.85  5.101
## 3   AO  50   50   57.82  9.151
## 4   AO  55   55   20.05  3.208
## 5   TC  57   57   25.72  3.906
## 6   AO  83   83   28.73  4.599
```

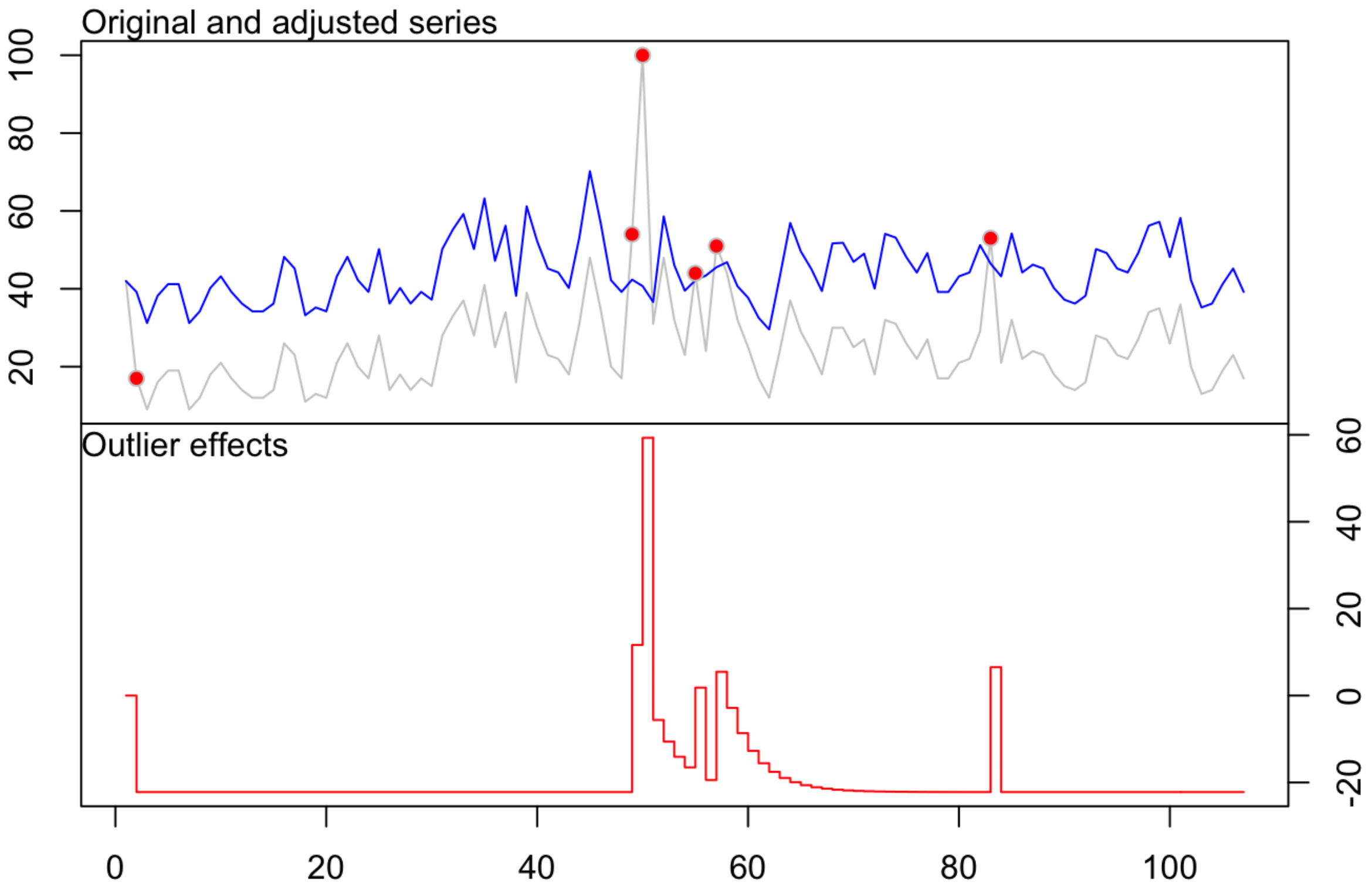

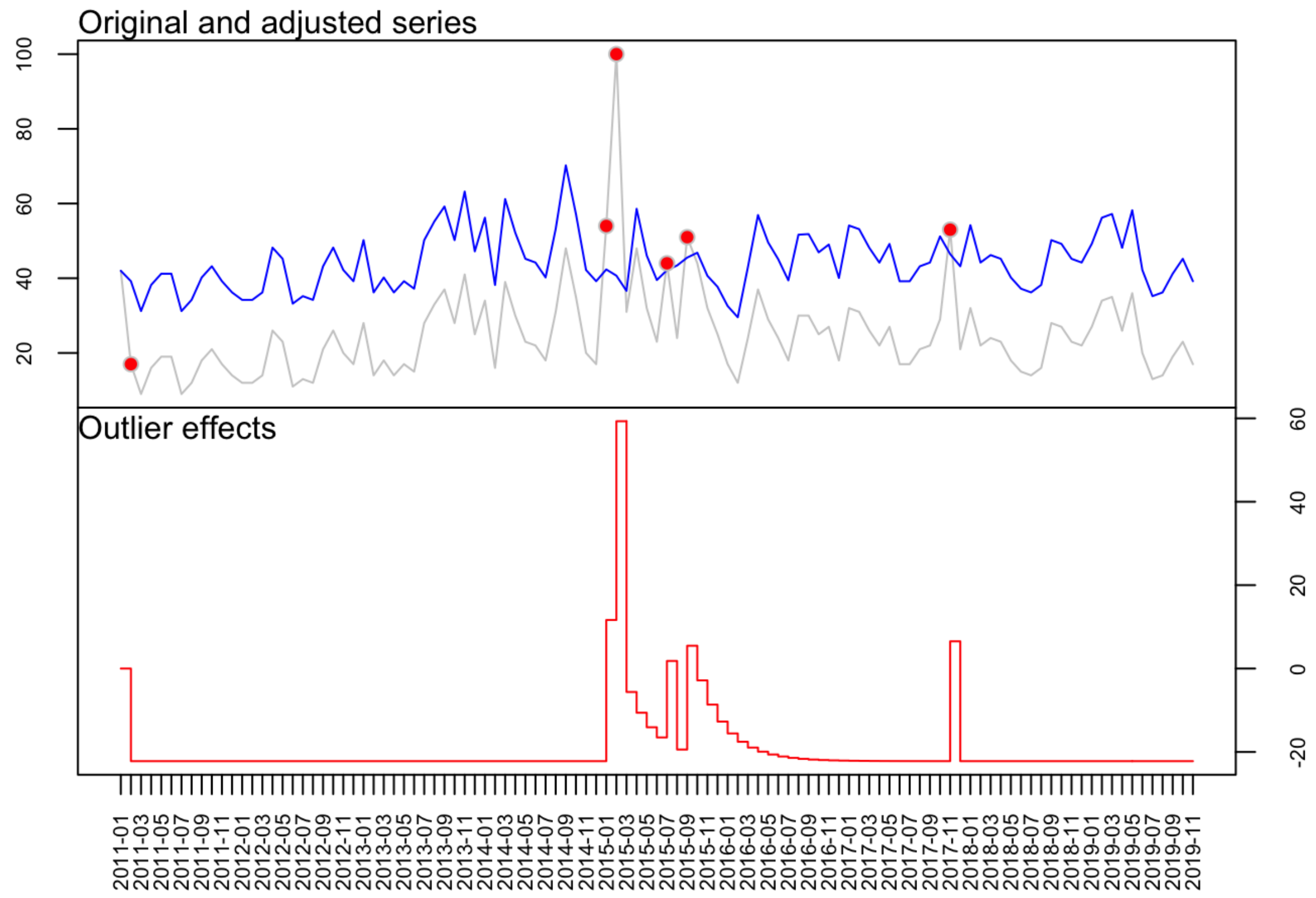

France (risks)

# General

```
## Series: fr_risk_gen.ts
## Regression with ARIMA(0,0,1) errors
##
## Coefficients:
##          ma1  intercept     TC82    AO95    LS106
##       0.4252    39.8543  40.1904  41.223  37.9951
## s.e.  0.0820     1.6625  10.5248  11.316  10.2745
##
## sigma^2 estimated as 144.1:  log likelihood=-415.3
## AIC=842.6   AICc=843.44   BIC=858.63
##
## Outliers:
##   type ind time coefhat tstat
## 1   TC  82   82   40.19 3.819
## 2   AO  95   95   41.22 3.643
## 3   LS 106  106   38.00 3.698
```

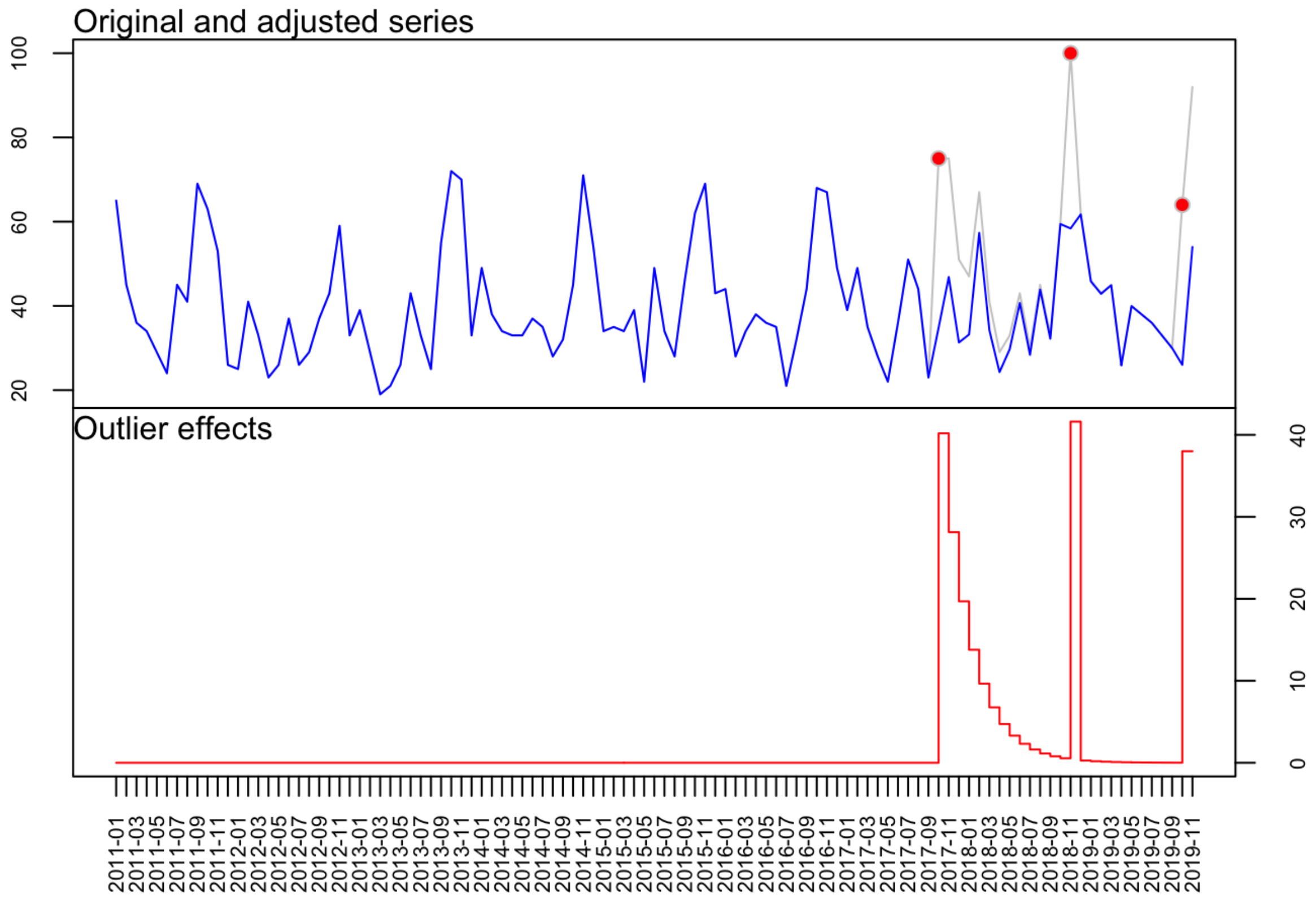

# Specific

```
## Series: fr_risk_spec.ts
## Regression with ARIMA(0,1,1) errors
##
## Coefficients:
##          ma1      AO2     TC13     TC53     TC69     TC79      AO80     TC81
##      -0.7875  16.4404  21.7538  14.6856  14.4771  40.9246  -19.4497  27.1683
```

```
## s.e.   0.0622   4.8123   4.1563   4.0530   4.0819   4.7630    5.4215   4.4201
##
## sigma^2 estimated as 26.02:  log likelihood=-319.46
## AIC=656.91   AICc=658.79   BIC=680.88
##
## Outliers:
##   type ind time coefhat  tstat
## 1   AO   2    2  16.44  3.416
## 2   TC  13   13   21.75  5.234
## 3   TC  53   53   14.69  3.623
## 4   TC  69   69   14.48  3.547
## 5   TC  79   79   40.92  8.592
## 6   AO  80   80  -19.45 -3.587
## 7   TC  81   81   27.17  6.146
```

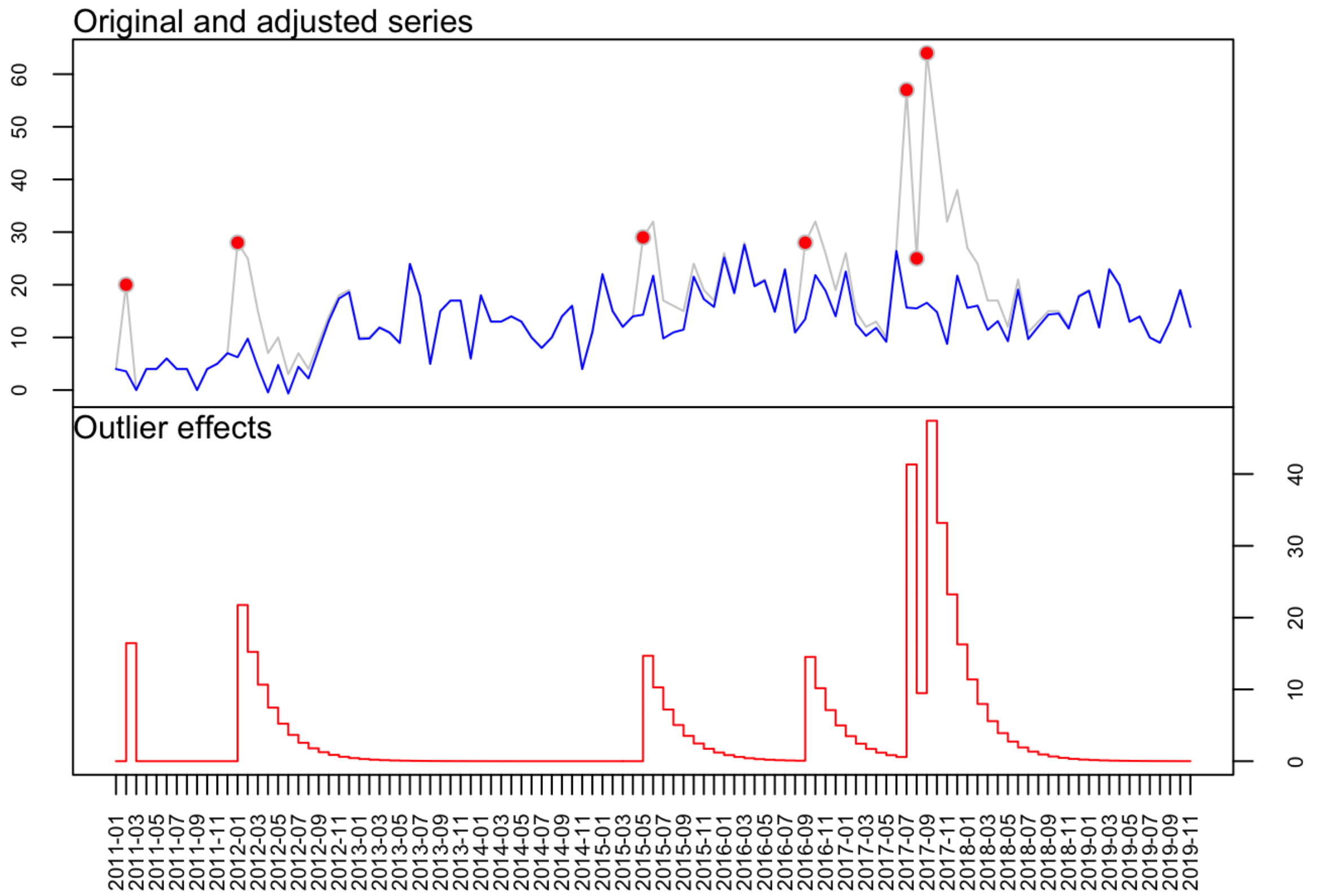

# Italy

## General

```
## Series: ita_risk_gen.ts
## Regression with ARIMA(0,1,1) errors
##
## Coefficients:
##          ma1     AO62     TC71     TC73     AO95     AO107
##      -0.7331  32.4526  29.1050  32.3329  62.8628  51.4769
## s.e.   0.0678   8.4019   7.7347   7.7284   8.3844   9.0000
```

```
##
## sigma^2 estimated as 85.82:  log likelihood=-383.67
## AIC=781.35   AICc=782.49   BIC=799.99
##
## Outliers:
##   type ind time coefhat tstat
## 1   AO  62   62   32.45 3.863
## 2   TC  71   71   29.11 3.763
## 3   TC  73   73   32.33 4.184
## 4   AO  95   95   62.86 7.498
## 5   AO 107  107   51.48 5.720
```

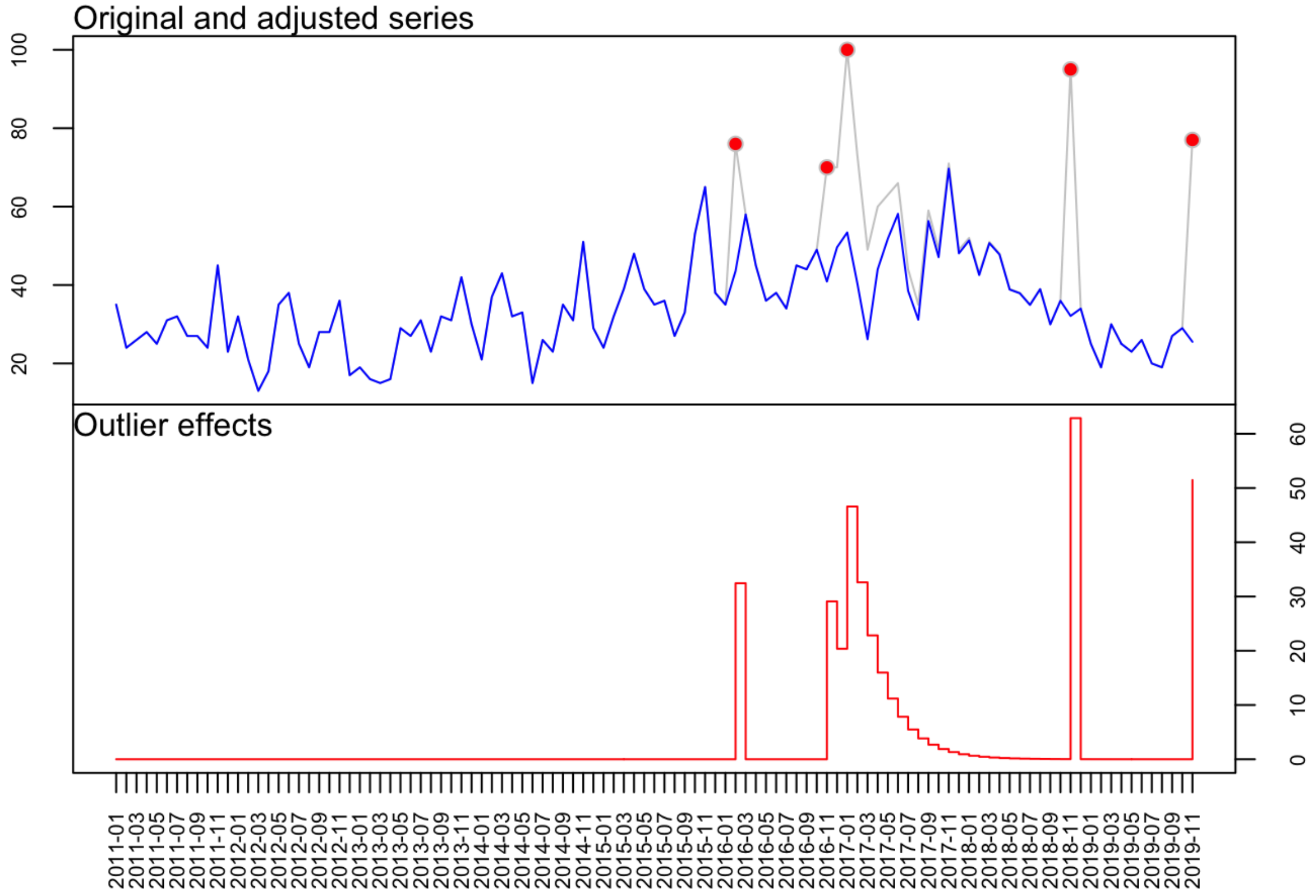

# Specific

```
## Series: ita_risk_spec.ts
## Regression with ARIMA(0,1,1) errors
##
## Coefficients:
##           ma1     AO39     AO47     AO58     TC77     AO81
##       -0.6776  23.9915  57.7996  30.1081  32.2230  27.1764
## s.e.   0.0749   6.6603   6.6446   6.6402   6.8522   6.7003
##
## sigma^2 estimated as 55.71:  log likelihood=-360.7
## AIC=735.4   AICc=736.54   BIC=754.04
##
## Outliers:
##   type ind time coefhat tstat
```

## 1   AO  39    39    23.99 3.602
## 2   AO  47    47    57.80 8.699
## 3   AO  58    58    30.11 4.534
## 4   TC  77    77    32.22 4.703
## 5   AO  81    81    27.18 4.056

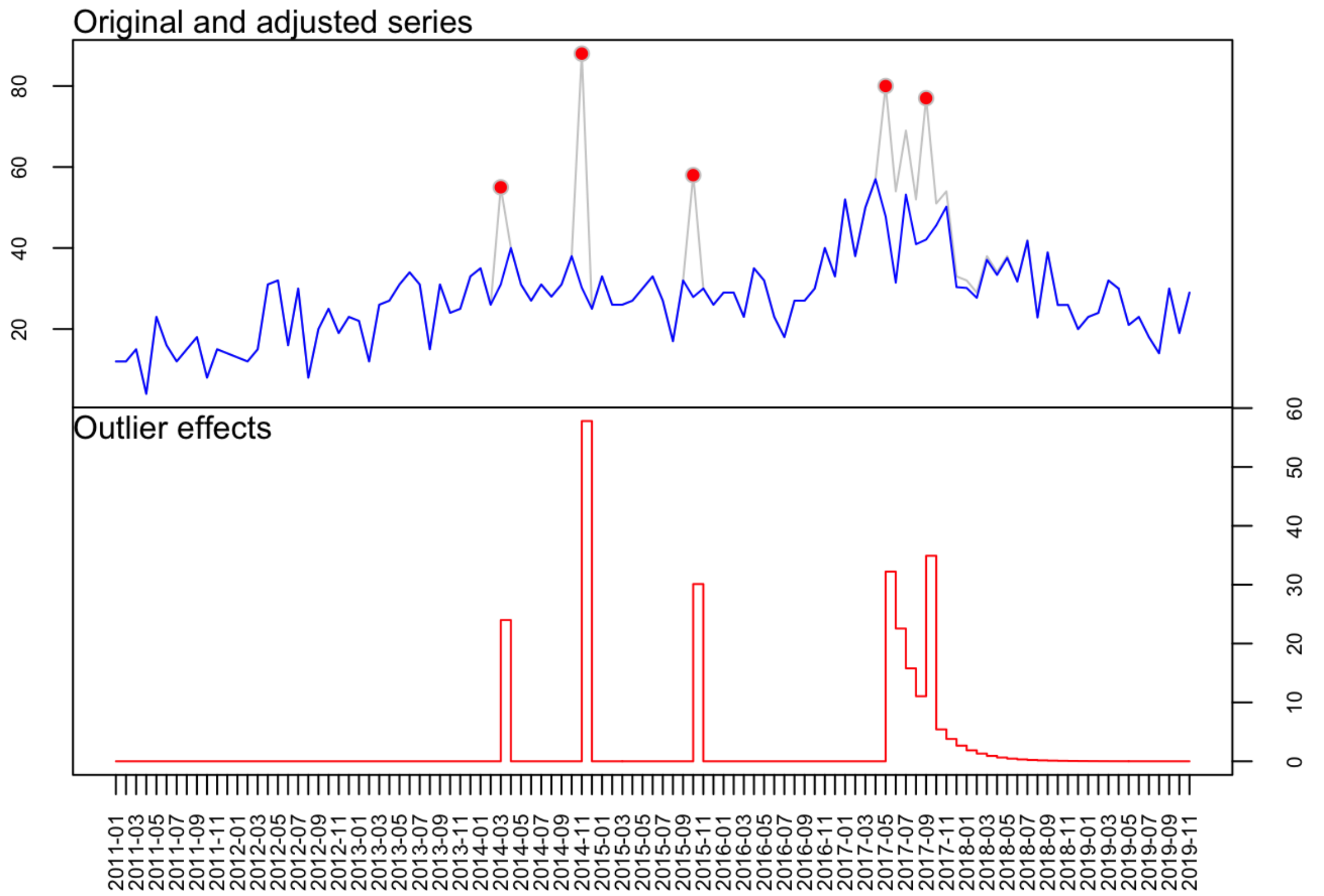

###Serbia (risks)

## General

## Series:
## ARIMA(0,1,1)
##
## Coefficients:
##           ma1
##       -0.8475
## s.e.   0.0667
##
## sigma^2 estimated as 374:  log likelihood=-464.53
## AIC=933.05   AICc=933.17   BIC=938.38
##
## No outliers were detected.

## 'x' does not contain outliers to display

## NULL

## Specific

```
## Series:
## ARIMA(0,0,0) with non-zero mean
##
## Coefficients:
## intercept
##         0
##
## sigma^2 estimated as 0:  log likelihood=Inf
## AIC=-Inf   AICc=-Inf   BIC=-Inf
##
## No outliers were detected.
```

```
## 'x' does not contain outliers to display
```

```
## NULL
```

# Washington (risks)

## General

```
## Series: wash_risk_gen.ts
## Regression with ARIMA(0,0,0) errors
##
## Coefficients:
##       intercept     AO94     LS98
##         22.9583  31.0417  16.6417
## s.e.     0.7348   7.2370   2.3923
##
## sigma^2 estimated as 53.33:  log likelihood=-363.05
## AIC=734.09   AICc=734.49   BIC=744.79
##
## Outliers:
##   type ind time coefhat tstat
## 1   AO  94   94   31.04 4.289
## 2   LS  98   98   16.64 6.956
```

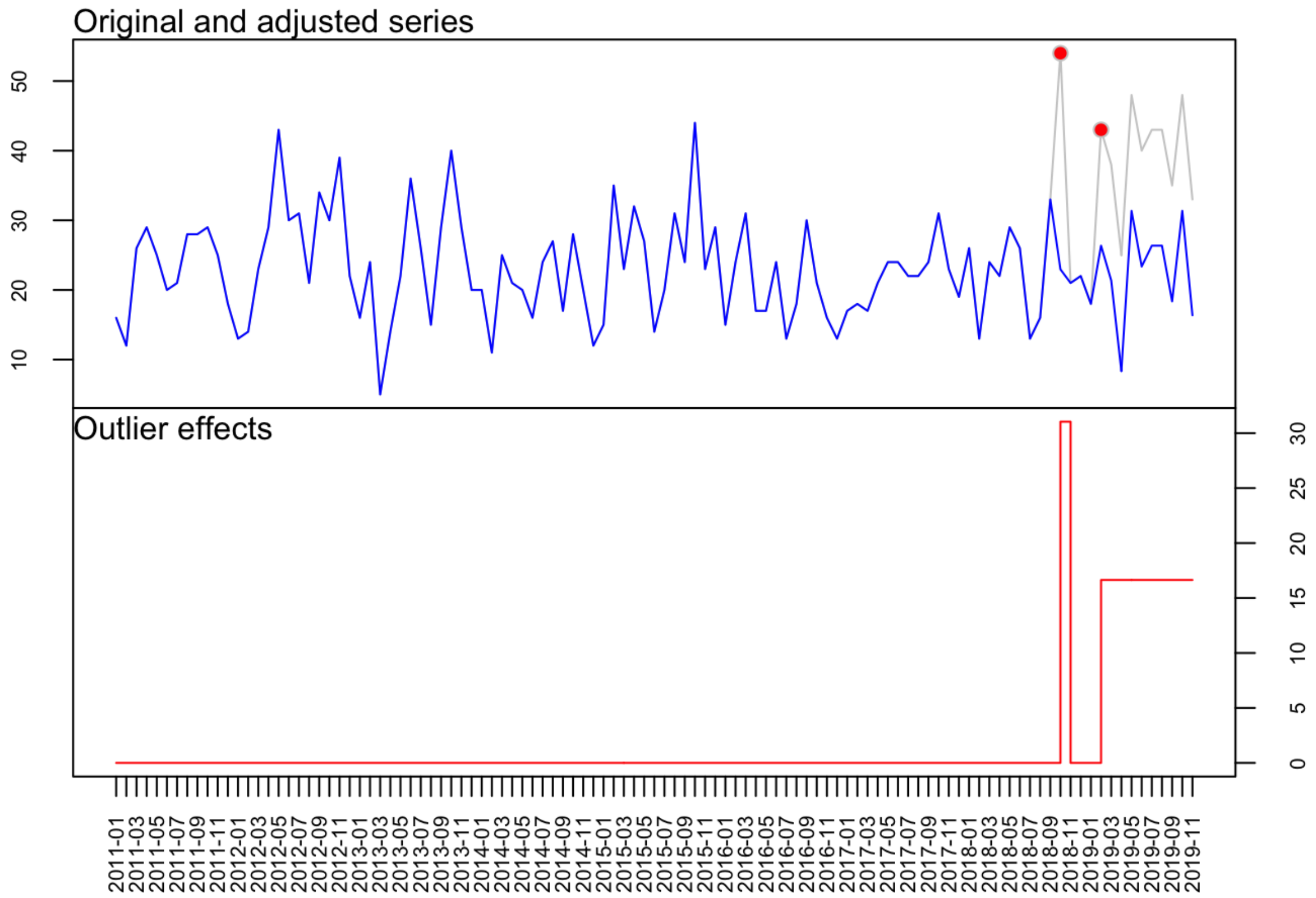

## Specific

```
## Series: wash_risk_spec.ts
## Regression with ARIMA(0,0,0) errors
##
## Coefficients:
##       intercept     AO50     AO58     AO83     TC97
##         19.4110  80.5890  26.5890  29.5890  20.2787
## s.e.     0.7747   7.7298   7.7298   7.7298   5.6432
##
## sigma^2 estimated as 62.05:  log likelihood=-370.11
## AIC=752.22   AICc=753.06   BIC=768.26
##
## Outliers:
##   type ind time coefhat  tstat
## 1   AO  50   50   80.59 10.426
## 2   AO  58   58   26.59  3.440
## 3   AO  83   83   29.59  3.828
## 4   TC  97   97   20.28  3.593
```

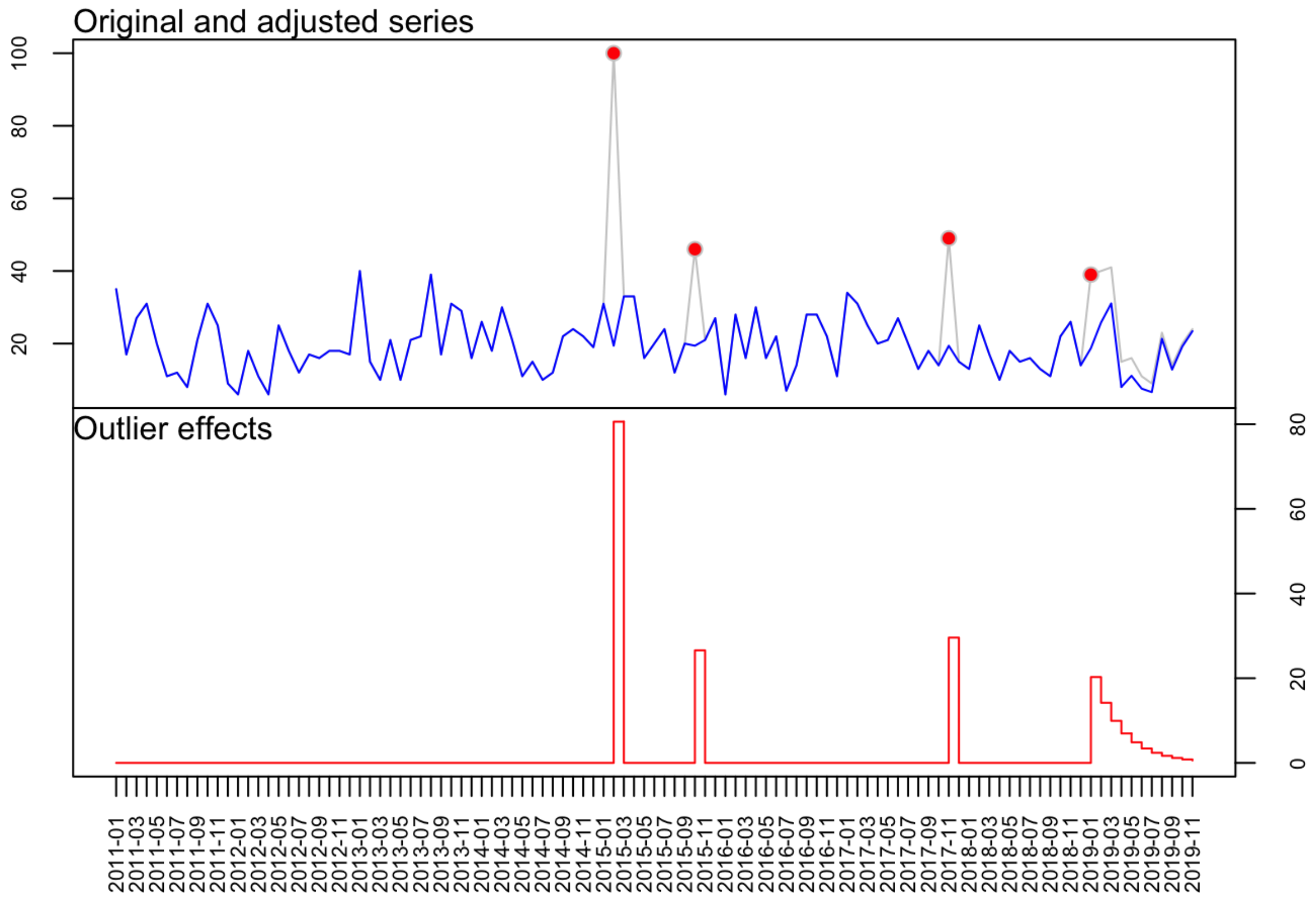
Original and adjusted series
Outlier effects